\documentclass[a4paper,11pt]{article}
 \pdfoutput=1
\usepackage{graphicx,epsfig}
\usepackage{fancyhdr,fancybox}
\usepackage{indentfirst}
\usepackage{verbatim}
\usepackage[sort&compress, numbers]{natbib}
\usepackage{geometry}
\usepackage{extarrows,chemarrow,xypic}
\usepackage[small]{caption2}
\usepackage{color}
\usepackage{microtype}
\DisableLigatures[f]{encoding = *, family = *}

\usepackage{times}

\usepackage{times}
\usepackage{amssymb}
\usepackage{amsthm}
\usepackage{bbm}
\usepackage{hyperref}

\newcommand{\paperfont}{\fontsize{11pt}{1.2\baselineskip}\selectfont}
\begin{document}

\theoremstyle{definition}
\makeatletter
\thm@headfont{\bf}
\makeatother
\newtheorem{definition}{Definition}
\newtheorem{example}{Example}
\newtheorem{theorem}{Theorem}
\newtheorem{lemma}{Lemma}
\newtheorem{corollary}{Corollary}
\newtheorem{remark}{Remark}
\newtheorem{proposition}{Proposition}

\lhead{}
\rhead{}
\lfoot{}
\rfoot{}

\renewcommand{\refname}{References}
\renewcommand{\figurename}{Fig.}
\renewcommand{\tablename}{Table}
\renewcommand{\proofname}{Proof}

\newcommand{\diag}{\textrm{diag}}
\newcommand{\one}{\mathbbm{1}}
\newcommand{\hyper}{{}_2F_1}
\newcommand{\confluent}{{}_1F_1}


\title{\textbf{Small protein number effects in stochastic models of autoregulated bursty gene expression}}
\author{Chen Jia$^{1}$,\;\;\;Ramon Grima$^{2,*}$ \\
\footnotesize $^1$Division of Applied and Computational Mathematics, Beijing Computational Science Research Center, Beijing 100193, China \\
\footnotesize $^2$ School of Biological Sciences, University of Edinburgh, U.K. \\
\footnotesize $^*$ Correspondence: Ramon.Grima@ed.ac.uk}
\date{}                              
\maketitle                           
\thispagestyle{empty}                

\paperfont

\begin{abstract}
A stochastic model of autoregulated bursty gene expression by Kumar et al. [Phys. Rev. Lett. 113, 268105 (2014)] has been exactly solved in steady-state conditions under the implicit assumption that protein numbers are sufficiently large such that fluctuations in protein numbers due to reversible protein-promoter binding can be ignored. Here we derive an alternative model that takes into account these fluctuations and hence can be used to study low protein number effects. The exact steady-state protein number distributions is derived as a sum of Gaussian hypergeometric functions. We use the theory to study how promoter switching rates and the type of feedback influence the size of protein noise and noise-induced bistability. Furthermore we show that our model predictions for the protein number distribution are significantly different from those of Kumar et al. when the protein mean is small, gene switching is fast, and protein binding is faster than unbinding.   
%
\end{abstract}

\section{Introduction}
One of the most common gene network motifs is an autoregulatory feedback loop whereby protein expressed from a gene activates or represses its own expression \cite{shen2002network}. It has been estimated that 40\% of all transcription factors in \emph{Escherichia coli} self-regulate \cite{rosenfeld2002negative} with most of them participating in negative autoregulation \cite{shen2002network}. Feedback leads to a regulation of the magnitude of intrinsic noise \cite{simpson2003frequency, hasty2000noise,liu2016decomposition, jia2017stochastic}, and also to changes in the response time and relaxation time of transcription networks \cite{rosenfeld2002negative, jia2018relaxation}.

The predictions by stochastic models of autoregulation have also been shown to lead to considerably different dynamics than those by deterministic models. For example, while deterministic models of non-cooperative autoregulation predict monostable gene expression for all parameter values, stochastic models of the same set of reactions show switching between two distinct gene expression levels and thus lead to bistability \cite{to2010noise, grima2012steady}. It has also been shown that in the presence of noise, feedback loops can lead to sustained oscillations in protein numbers in regions of the parameter space where deterministic models predict damped oscillations \cite{ko2010emergence, thomas2012slow, thomas2013signatures}. These results suggest that stochastic models are necessary to understand the intracellular dynamics of biological systems utilizing a combination of positive and negative feedback loops and in which at least one molecular component is present in low copy numbers, e.g. circadian clocks \cite{forger2005stochastic, guerriero2011stochastic}.

The stochastic properties of an autoregulatory gene circuit have been explored mostly by stochastic simulations and to a lesser extent by analytical solutions of various discrete, continuous, and hybrid gene expression models (see \cite{holehouse2019stochastic} for a recent review). Discrete models are those in which gene, mRNA, and protein numbers change by discrete integer amounts when reactions occur; in continuous models, fluctuations correspond to hops on the real axis rather than on the integer axis \cite{friedman2006linking, thomas2013reliable, mackey2013dynamic, bokes2015protein}; in hybrid models, some fluctuations are modeled discretely (such as those of genes) while other types of fluctuations (such as those of mRNAs and proteins) are modeled in a continuous sense \cite{thomas2014phenotypic, ge2015stochastic, jia2017emergent, jia2019single}. Here we focus on discrete models since these are the most realistic among the three types (continuous and hybrid models have the advantage of possessing simpler distribution solutions that can be easier to interpret).

In the literature, there are two exact solutions for the steady-state protein number distribution of stochastic autoregulatory models \cite{grima2012steady,kumar2014exact}. Both models assume that a single gene can exist in one of two states (with two different protein production rates) and that reversible binding of a protein molecule to a gene leads to switching from one gene state to the other. The differences between the two models are as follows. The model solved by Grima et al. \cite{grima2012steady} (henceforth referred to as the Grima model; see Fig. \ref{model}(a) for an illustration) assumes that (i) protein molecules are produced one at a time and (ii) when a protein molecule binds to a gene, the protein copy number decreases by one whereas it increases by one when unbinding occurs. In contrast, the model solved by Kumar et al. \cite{kumar2014exact} (henceforth referred to as the Kumar model; see Fig. \ref{model}(b) for an illustration) assumes that (i) proteins are produced in random bursts whose size is a random number sampled from a geometric distribution and (ii) when a protein molecule binds to a gene or unbinds from it, there is no change in the protein copy number. The advantage of the Kumar model is its modeling of bursty protein expression which is in agreement with experimental data \cite{yu2006probing}; its disadvantage (unlike the Grima model) is the implicit neglect of fluctuations due to reversible protein-promoter binding which necessarily implies that it cannot precisely capture low protein number effects. Taking into account the latter is important because the copy number of DNA-binding proteins can be in the single digits \cite{taniguchi2010quantifying}.

In this paper, we remedy the shortcomings of the Grima and Kumar models, by deriving and solving a discrete stochastic model of autoregulatory feedback loops that takes into account both translational bursting and protein number fluctuations during the binding-unbinding process. The paper is organized as follows. In Section 2, starting from a stochastic model of autoregulation describing both mRNA and protein dynamics (henceforth referred to as the full model; see Fig. \ref{model}(c) for an illustration), we use multiscale decimation theory to derive a reduced stochastic model of autoregulation for the protein dynamics in the limit of fast mRNA degradation. The chemical master equation (CME) of this reduced model is similar to that of the Kumar model except that the propensities include protein fluctuations during the binding-unbinding process --- we henceforth refer to this as the modified Kumar model which is illustrated in Fig. \ref{model}(d). In Section 3, we provide an exact analytical solution for the steady-state protein distribution of the modified Kumar model in terms of Gaussian hypergeometric functions, show its simplification in the cases of fast and slow gene switching, discuss the influence of feedback on protein noise and bistability, and verify the theory using stochastic simulations. In Section 4, we show that the modified Kumar model reduces to the Grima model under certain conditions. In Section 5, we compare the modified Kumar model with the Kumar model, showing agreement between the two under slow gene switching and disagreement under fast gene switching and strong protein-promoter interactions. We also investigate how the relative sensitivity of protein noise to model parameters differs between the two models. We conclude in Section 6.

\section{Derivation of the modified Kumar model from a fine-grained model}
\begin{figure}[!htb]
\centerline{\includegraphics[width=1.0\textwidth]{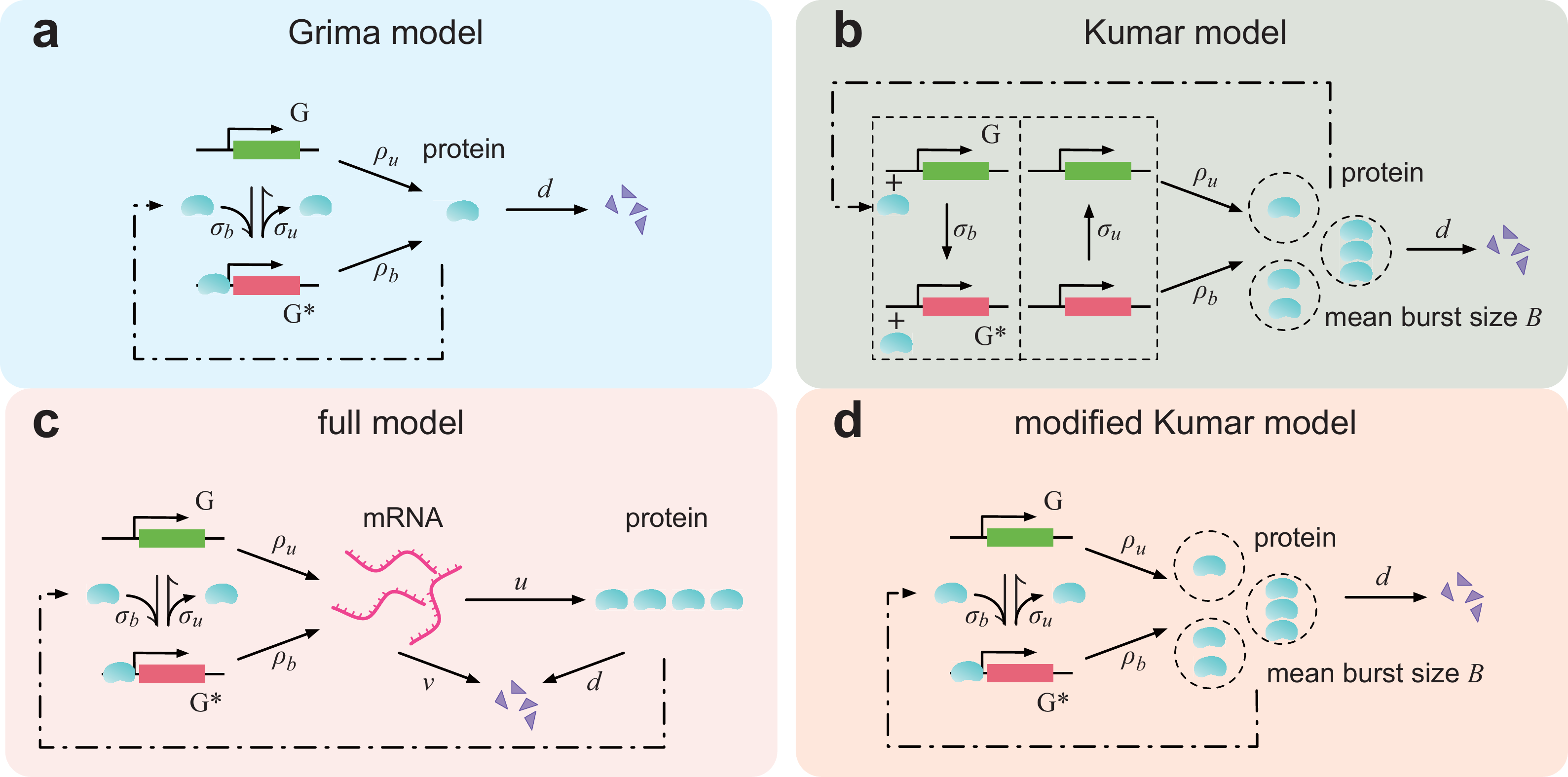}}
\caption{\textbf{Discrete models of an autoregulatory gene network.} (a) Grima model \cite{grima2012steady}. This model describes protein dynamics only. It neglects translational bursting but takes into account fluctuations in protein number due to binding to and unbinding from the gene. (b) Kumar model \cite{kumar2014exact}. This model describes protein dynamics only. It takes into account translational bursting but neglects protein fluctuations during the binding-unbinding process, i.e. when a protein molecule binds to the promoter, there is no change in the protein number. There are two dotted boxes in this figure; the left one displays the reaction $G+P \xlongrightarrow{\sigma_b} G^*+P$ and the right one displays the reaction $G^*\xlongrightarrow{\sigma_u} G$.
(c) Fine-grained full model. This model has both mRNA and protein descriptions. It (implicitly) takes into account translational bursting (via the mRNA description) and also fluctuations in protein number during the binding-unbinding process. (d) Modified Kumar model. This model, which is derived from the full model in the limit of fast mRNA degradation, takes into account both translational bursting and protein fluctuations during the binding-unbinding process.}\label{model}
\end{figure}

Based on the central dogma of molecular biology, the gene expression kinetics for an autoregulatory gene network in an individual cell has a standard three-stage representation involving gene switching, transcription, and translation (Fig. \ref{model}(c)) \cite{shahrezaei2008analytical}. Due to autoregulation, the promoter could either be free or bound to a protein molecule to form a promoter-protein complex. Let $G$ and $G^*$ denote the unbound and bound states of the gene, respectively, let $M$ be the corresponding mRNA, and let $P$ denote the corresponding protein. Then the effective reactions describing the autoregulatory gene circuit are given by:
\begin{gather*}
G+P \xlongrightarrow{\sigma_b} G^*,\;\;\;G^*\xlongrightarrow{\sigma_u} G+P,\\
G\xlongrightarrow{\rho_u} G+M,\;\;\;G^*\xlongrightarrow{\rho_b} G^*+M,\\
M\xlongrightarrow{u} M+P,\;\;\;M \xlongrightarrow{v} \varnothing,\;\;\;
P \xlongrightarrow{d} \varnothing.
\end{gather*}
Here $\sigma_b$ is the binding rate of protein to the promoter which characterizes the strength of feedback; $\sigma_u$ is the unbinding rate of protein from the promoter; $\rho_u$ and $\rho_b$ are the transcription rates when the gene is unbound and bound to protein, respectively; $u$ is the translation rate; $v$ and $d$ are the degradation rates of mRNA and protein, respectively. The reaction scheme describes a positive feedback loop if $\rho_b > \rho_u$ and describes a negative feedback loop if $\rho_b < \rho_u$. We shall refer to this fine-grained model as the full model.

\begin{figure}[!htb]
\centerline{\includegraphics[width=0.8\textwidth]{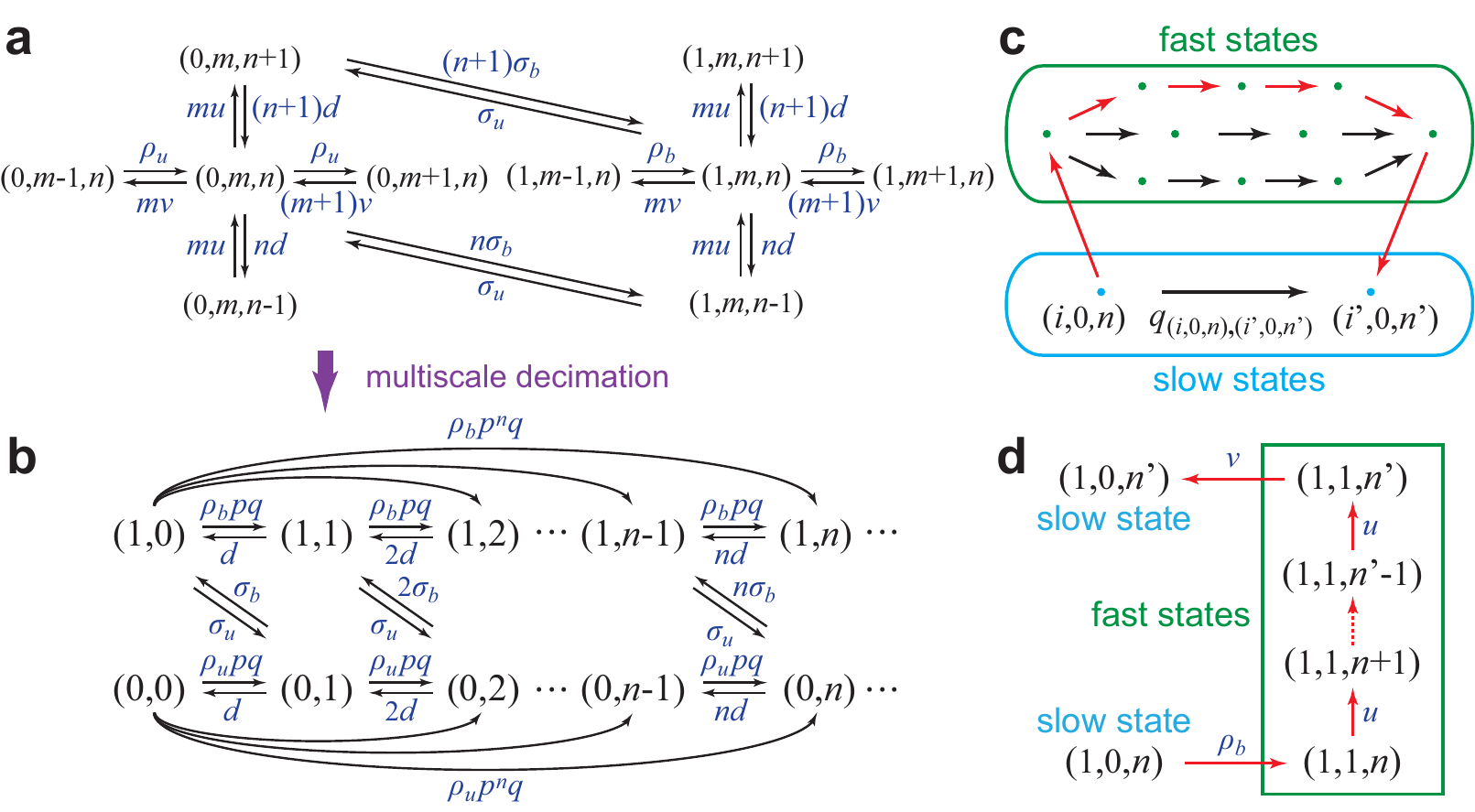}}
\caption{\textbf{Full and reduced Markovian models of autoregulatory gene networks.} (a) Transition diagram of the full Markovian model, where the microstate of the gene is described by an ordered triple $(i,m,n)$. (b) Transition diagram of the reduced Markovian model when mRNA decays much faster than protein, where the microstate of the gene is described by an ordered pair $(i,n)$. (c) Schematic diagram of the decimation method of multiscale model simplification. The effective transition rate from microstate $(i,0,n)$ to microstate $(i',0,n')$ is the sum of the direct transition rate and the contribution of indirect transitions via all fast transition paths. (d) A typical fast transition path of the full model.}\label{simplification}
\end{figure}

The microstate of the gene of interest can be represented by an ordered triple $(i,m,n)$: the gene state $i$ with $i = 0,1$ corresponding to the unbound and bound states, respectively, the mRNA copy number $m$, and the protein copy number $n$. Let $p_{i,m,n}$ denote the probability of having $m$ copies of mRNA and $n$ copies of protein when the gene is in state $i$. Then the stochastic gene expression kinetics can be described by the Markov jump process illustrated in Fig. \ref{simplification}(a). The evolution of the Markovian model is governed by the CME
\begin{equation*}\left\{
\begin{split}
\dot p_{0,m,n} =&\; \rho_up_{0,m-1,n}+(m+1)vp_{0,m+1,n}+mup_{0,m,n-1}\\
&\; +(n+1)dp_{0,m,n+1}+\sigma_up_{1,m,n-1}\\
&\; -(\rho_u+mv+mu+nd+n\sigma_b)p_{0,m,n},\\
\dot p_{1,m,n} =&\; \rho_bp_{1,m-1,n}+(m+1)vp_{1,m+1,n}+mup_{1,m,n-1}\\
&\; +(n+1)dp_{1,m,n+1}+(n+1)\sigma_bp_{0,m,n+1}\\
&\; -(\rho_b+mv+mu+nd+\sigma_u)p_{1,m,n}.
\end{split}\right.
\end{equation*}

Experimentally, it is commonly observed that mRNA decays much faster than protein. For example, mRNA lifetimes in prokaryotes are usually of the order of a few minutes, whereas protein lifetimes are generally on the order of tens of minutes to many hours \cite{bernstein2002global}. Due to timescale separation of the underlying biochemical reaction kinetics, the full model can be reduced to a simpler one. The model reduction technique described below is similar to but slightly different from the one described in \cite{jia2017simplification}, where the author considered a different full model which neglects fluctuations in protein number during the binding-unbinding process.

Specifically, let $\lambda = v/d$ denote the ratio of the degradation rates of mRNA and protein. Here we make the classical assumption that $\lambda\gg 1$ and $u/v$ is strictly positive and bounded \cite{shahrezaei2008analytical}. In addition, let $q_{(i,m.n),(i',m'.n')}$ denote the transition rate of the Markovian model from microstate $(i,m,n)$ to microstate $(i',m',n')$ and let
\begin{equation*}
q_{(i,m,n)} = \sum_{(i',m',n')\neq(i,m,n)}q_{(i,m,n),(i',m',n')}
\end{equation*}
denote the rate at which the system leaves microstate $(i,m,n)$, which is defined as the sum of transition rates from $(i,m,n)$ to other microstates. Since $\lambda\gg 1$, we say that $(i,m,n)$ is a fast state if
\begin{equation*}
\lim_{\lambda\rightarrow\infty}q_{(i,m,n)} = \infty,
\end{equation*}
and we say that $(i,m,n)$ is a slow state if
\begin{equation*}
\lim_{\lambda\rightarrow\infty}q_{(i,m,n)} < \infty.
\end{equation*}
If $(i,m,n)$ is a fast state, then the time that the system stays in this state will be very short. Note that the limit of $\lambda \rightarrow \infty$ is taken at constant $u/v$ and $d$. It is easy to check that the leaving rates of all microstates are given by
\begin{gather*}
q_{(0,m,n)} = \rho_u+mv+mu+nd+n\sigma_b = m\lambda d(1+u/v)+nd+n\sigma_b+\rho_u, \\
q_{(1,m,n)} = \rho_b+mv+mu+nd+\sigma_u = m\lambda d(1+u/v)+nd+\sigma_u+\rho_b,
\end{gather*}
which imply that
\begin{equation*}
\lim_{\lambda\rightarrow\infty}q_{(i,m,n)}
\begin{cases}
= \infty, &\textrm{if\;} m\geq 1,\\
< \infty, &\textrm{if\;} m = 0.
\end{cases}
\end{equation*}
Therefore, all microstates $(i,m,n)$ for $m\geq 1$ are fast states and all microstates $(i,0,n)$ are slow states. By using a classical simplification method of multiscale Markov jump processes called decimation \cite{pigolotti2008coarse, jia2016reduction, jia2016simplification, bo2016multiple, jia2017simplification}, the full Markovian model can be simplified to a reduced one by removal of all fast states. For simplicity, microstate $(i,0,n)$ will be denoted by $(i,n)$ in the reduced model.

The remaining question is to determine the transition diagram and effective transition rates of the reduced Markovian model. This process is described as follows. Suppose that the full model jumps from microstate $(i,m,n)$ to another microstate at a particular time. When $\lambda\gg 1$, the transition probability from $(i,m,n)$ to another microstate $(i',m',n')$ is given by
\begin{equation*}
w_{(i,m,n),(i',m',n')} = \lim_{\lambda\rightarrow\infty}\frac{q_{(i,m,n),(i',m',n')}}{q_{(i,m,n)}}.
\end{equation*}
Let $(i_1,m_1,n_1),\cdots,(i_k,m_k,n_k)$ be a sequence of microstates. We say that
\begin{equation*}
c: (i,m,n)\rightarrow(i_1,m_1,n_1)\rightarrow\cdots\rightarrow(i_k,m_k,n_k)\rightarrow(i',m',n')
\end{equation*}
is a fast transition path from $(i,m,n)$ to $(i',m',n')$ if the intermediate microstates
\begin{equation*}
(i_1,m_1,n_1),\cdots,(i_k,m_k,n_k)
\end{equation*}
are all fast states. Moreover, the probability weight $w_c$ of the fast transition path $c$ is defined as
\begin{equation*}
w_c = q_{(i,m,n),(i_1,m_1,n_1)}w_{(i_1,m_1,n_1),(i_2,m_2,n_2)}\cdots w_{(i_k,m_k,n_k),(i',m',n')}.
\end{equation*}
According to decimation theory \cite{pigolotti2008coarse, jia2016reduction, jia2016simplification, bo2016multiple, jia2017simplification}, the effective transition rate $\tilde{q}_{(i,n),(i',n')}$ from microstate $(i,n)$ to microstate $(i',n')$ is given by
\begin{equation*}
\tilde{q}_{(i,n),(i',n')} = q_{(i,0,n),(i',0,n')}+\sum_cw_c,
\end{equation*}
where $c$ ranges over all fast transition paths from $(i,0,n)$ to $(i',0,n')$. This formula indicates that the effective transition rate from $(i,n)$ to $(i',n')$ is the superposition of two parts: the direct transition rate and the contribution of indirect transitions via all fast transition paths, as depicted in Fig. \ref{simplification}(c).

Since the intermediate microstates of a fast transition path $c$ are all fast states, in order for the path to have a positive probability weight, all the intermediate transitions along this path must satisfy
\begin{equation}\label{criterion}
\lim_{\lambda\rightarrow\infty}q_{(i_1,m_1,n_1),(i_2,m_2,n_2)} = \cdots = \lim_{\lambda\rightarrow\infty}q_{(i_k,m_k,n_k),(i',m',n')} = \infty.
\end{equation}
By using this criterion, it is easy to see that the full model has only one type of fast transition paths with positive probability weight, which is given by (see the red arrows in Fig. \ref{simplification}(d))
\begin{equation}\label{path}
(i,0,n)\rightarrow(i,1,n)\rightarrow(i,1,n+1)\rightarrow\cdots\rightarrow(i,1,n')\rightarrow(i,0,n'),\;\;\;n'>n.
\end{equation}
This is because any fast transition path from $(i,0,n)$ to $(i,0,n')$ with positive probability weight cannot pass through microstate $(i,m,k)$ for some $m\geq 2$ and $k\geq 0$. Otherwise, there must be an intermediate transition along the path from $(i,m-1,k)$ to $(i,m,k)$ with transition rate being $\rho_u$ or $\rho_b$, which does not diverge as $\lambda\rightarrow\infty$. This contradicts the criterion \eqref{criterion}. Moreover, since the intermediate transition rates along the path (2) are all given by $u$ or $v$, and $u,v\rightarrow\infty$ as $\lambda\rightarrow\infty$, it follows that the path (2) satisfies the criterion (1).

To proceed, we define two constants $p$ and $q$ as
\begin{equation*}
p = \frac{u}{u+v},\;\;\;q = \frac{v}{u+v}.
\end{equation*}
Since $\lambda\gg 1$, the transition probabilities along the path \eqref{path} are given by
\begin{gather*}
w_{(i,1,n),(i,1,n+1)} = \lim_{\lambda\rightarrow\infty}\frac{u}{q_{(i,1,n)}} = p,\;\;\;
w_{(i,1,n),(i,0,n)} = \lim_{\lambda\rightarrow\infty}\frac{v}{q_{(i,1,n)}} = q.
\end{gather*}
Therefore, the probability weight of this path is $\rho_up^{n'-n}q$ if $i = 0$ and is $\rho_bp^{n'-n}q$ if $i = 1$. Since there is no direct transition, the effective transition rate from $(i,n)$ to $(i,n')$ with $n'>n$ is exactly the indirect transition rate via the fast transition path \eqref{path}
\begin{equation*}
\tilde{q}_{(i,n),(i,n')} = \begin{cases}
\rho_up^{n'-n}q,\;\;\;i = 0,\\
\rho_bp^{n'-n}q,\;\;\;i = 1.
\end{cases}
\end{equation*}
This formula indicates that the reduced model may produce large jumps of protein number within a very short period, which correspond to random translational bursts. Each random burst corresponds to a fast transition path of the full Markovian model. The above computations can be understood intuitively as follows. Since $v\gg d$ and $u/v$ is finite, the process of protein synthesis followed by mRNA degradation is essentially instantaneous. Once a transcript is synthesized, it can either produce a protein molecule with probability $p = u/(u+v)$ or be degraded with probability $q = 1-p = v/(u+v)$. Thus, the probability that a transcript produces $k$ copies of protein before it is finally degraded will be $p^kq$, which follows a geometric distribution. Note that since the protein burst size is geometrically distributed, its expected value is given by
\begin{equation*}
B = \sum_{k=0}^\infty kp^kq = \frac{p}{q} = \frac{u}{v}.
\end{equation*}

So far, we have obtained the transition diagram and all effective transition rates of the reduced model, as depicted in Fig. \ref{simplification}(b). Let $p_{i,n}$ denote the probability of having $n$ copies of protein when the gene is in state $i$. Then the evolution of the reduced model is governed by the coupled set of master equations
\begin{equation}\label{masterreduced}\left\{
\begin{split}
\dot p_{0,n} &= \sum_{k=0}^{n-1}\rho_up^{n-k}qp_{0,k}+(n+1)dp_{0,n+1}+\sigma_up_{1,n-1}
-(\rho_up+nd+n\sigma_b)p_{0,n},\\
\dot p_{1,n} &= \sum_{k=0}^{n-1}\rho_bp^{n-k}qp_{1,k}+(n+1)dp_{1,n+1}+(n+1)\sigma_bp_{0,n+1}
-(\rho_bp+nd+\sigma_u)p_{1,n}.
\end{split}\right.
\end{equation}
This is exactly the CME of the following chemical reaction system (see Fig. \ref{model}(d) for an illustration):
\begin{align}\label{MK}
\nonumber &G+P \xlongrightarrow{\sigma_b} G^*,\;\;\;G^*\xlongrightarrow{\sigma_u} G+P,\\
&G\xlongrightarrow{\rho_up^kq} G+kP,\;\;\;G^*\xlongrightarrow{\rho_bp^kq} G^*+kP,\;\;\;k\geq 1,\\
\nonumber &P \xlongrightarrow{d} \varnothing.
\end{align}
This reaction scheme takes into account both translational bursting resulting from short-lived mRNA and protein fluctuations during the binding-unbinding process, i.e. when a protein molecule binds to the promoter, the protein number in a single cell is decreased by one and conversely it is increased by one when unbinding occurs. In fact, the reduced model modifies a model of autoregulatory gene networks proposed by Kumar et al. \cite{kumar2014exact}. In the Kumar model, the authors also take into account translational bursting but neglect protein fluctuations during the binding-unbinding process, i.e. when a protein molecule binds to or unbinds from the promoter, the protein number remains the same. In the following, we shall refer to the reduced model as the modified Kumar model. This model was first introduced (using intuitive arguments) and studied using approximation methods in \cite{cao2018linear}. A method to efficiently infer the parameters of this model from noisy experimental data by means of MCMC has been described in \cite{Cao2019accuracy}. 

\section{Exact analytic solution of the modified Kumar model}

\subsection{General case}
Here we derive the exact analytic solution for the steady-state protein number distribution of the modified Kumar model. For convenience, we normalize all model parameters by the protein degradation rate as
\begin{equation*}
r = \frac{\rho_u}{d},\;\;\;s = \frac{\rho_b}{d},\;\;\;
\mu = \frac{\sigma_b}{d},\;\;\;b = \frac{\sigma_u}{d}.
\end{equation*}
At steady-state, all probabilities are time-independent and thus Eq. \eqref{masterreduced} can be rewritten as
\begin{equation}\label{master}\left\{
\begin{split}
\sum_{k=0}^{n-1}rp^{n-k}qp_{0,k}+(n+1)p_{0,n+1}+bp_{1,n-1}-(rp+n+n\mu)p_{0,n} = 0,\\
\sum_{k=0}^{n-1}sp^{n-k}qp_{1,k}+(n+1)p_{1,n+1}+(n+1)\mu p_{0,n+1}-(sp+n+b)p_{1,n} = 0.
\end{split}\right.
\end{equation}
To proceed, we define a pair of generating functions
\begin{equation*}
f(z) = \sum_{n=0}^\infty p_{1,n}z^n,\;\;\;g(z) = \sum_{n=0}^\infty p_{0,n}z^n.
\end{equation*}
Let $p_n = p_{1,n}+p_{0,n}$ denote the probability of having $n$ copies of protein and let $F(z) = f(z)+g(z)$ denote its generating function. Then Eq. \eqref{master} can be converted into the following system of ordinary differential equations (ODEs):
\begin{gather}
\label{geneqss1} [sp(1-z)+b(1-pz)]f(z)-(1-z)(1-pz)f'(z)-\mu(1-pz)g'(z) = 0, \\
\label{geneqss2} rp(1-z)g(z)-[(1-z)-\mu z](1-pz)g'(z)-bz(1-pz)f(z) = 0.
\end{gather}
Solving these equations leads to (see Appendix A for details)
\begin{equation}\label{generating}
\begin{split}
f(z) =&\; \frac{r\mu pK}{(q+\mu)\beta}(1-pz)^{-s}\hyper(\alpha_1+1,\alpha_2+1;\beta+1;w(z-z_0)),\\
F(z) =&\; K(1-pz)^{-s}\bigg[\hyper(\alpha_1,\alpha_2;\beta;w(z-z_0)) \\
&\; +A(1-az)\hyper(\alpha_1+1,\alpha_2+1;\beta+1;w(z-z_0))\bigg],
\end{split}
\end{equation}
where $\hyper(\alpha_1,\alpha_2;\beta;z)$ is the Gaussian hypergeometric function,
\begin{gather}\label{modifiedkumar}
\alpha_1+\alpha_2 = \frac{b+r}{1+\mu}-s-1,\;\;\;
\alpha_1\alpha_2 = s+\frac{b(r-s)-r}{1+\mu},\;\;\;
\beta = \frac{b}{1+\mu}+\frac{r\mu p}{(1+\mu)(q+\mu)}, \\ \nonumber
A = \frac{(s-r+r\mu)p}{(q+\mu)\beta},\;\;\;a = \frac{s-r+s\mu}{s-r+r\mu},\;\;\;
w = \frac{(1+\mu)p}{q+\mu},\;\;\;z_0 = \frac{1}{1+\mu},
\end{gather}
and
\begin{equation*}
K = q^s[\hyper(\alpha_1,\alpha_2;\beta;w(1-z_0))+A(1-a)\hyper(\alpha_1+1,\alpha_2+1;\beta+1;w(1-z_0))]^{-1}
\end{equation*}
is a normalization constant that can be determined by solving $F(1) = 1$. Recall that the steady-state distribution of protein number can be obtained by taking the derivatives of the generating function $F(z)$ at zero:
\begin{equation*}
p_n = \frac{1}{n!}\frac{d^n}{dz^n}\Big|_{z=0}F(z).
\end{equation*}
Taking the derivatives of the generating function given by Eq. \eqref{generating} yields
{\footnotesize\begin{equation}
\label{pexactGaussian}
\begin{split}
p_n =&\; K\sum_{k=0}^n\frac{(\alpha_1)_k(\alpha_2)_k(s)_{n-k}}{(\beta)_k(1)_k(1)_{n-k}}
\hyper(\alpha_1+k,\alpha_2+k;\beta+k;-wz_0)w^kp^{n-k} \\
&\; +KA\sum_{k=0}^n\frac{(\alpha_1+1)_k(\alpha_2+1)_k(s)_{n-k}}{(\beta+1)_k(1)_k(1)_{n-k}}
\hyper(\alpha_1+1+k,\alpha_2+1+k;\beta+1+k;-wz_0)w^kp^{n-k} \\
&\; -KAa\sum_{k=0}^{n-1}\frac{(\alpha_1+1)_k(\alpha_2+1)_k(s)_{n-1-k}}{(\beta+1)_k(1)_k(1)_{n-1-k}}
\hyper(\alpha_1+1+k,\alpha_2+1+k;\beta+1+k;-wz_0)w^kp^{n-1-k}.
\end{split}
\end{equation}}

We next focus on two trivial special cases. In the case of $\sigma_u\gg\sigma_b,\rho_u\rho_b,d$, the gene is mostly in the unbound state and the parameters in Eq. \eqref{modifiedkumar} reduce to
\begin{equation*}
\alpha_1 = \beta = \frac{b}{1+\mu},\;\;\;\alpha_2 = r-s,\;\;\;A = 0.
\end{equation*}
Since $\alpha_1 = \beta$, we have (see Eq. 15.4.6 in \cite{special})
\begin{equation*}
\hyper(\alpha_1,\alpha_2;\beta;w(z-z_0)) = [1-w(z-z_0)]^{-\alpha_2} = \left(\frac{1+\mu}{q+\mu}\right)^{s-r}(1-pz)^{s-r},
\end{equation*}
and thus the generating function reduces to
\begin{equation*}
F(z) = K\left(\frac{1+\mu}{q+\mu}\right)^{s-r}(1-pz)^{-r}.
\end{equation*}
Therefore, the protein number distribution is negative binomial and given by \cite{paulsson2000random, shahrezaei2008analytical}
\begin{equation}\label{NB1}
p_n = \frac{(r)_n}{n!}p^nq^r,
\end{equation}
where $(x)_n = x(x+1)\cdots(x+n-1) = \Gamma(x+n)/\Gamma(x)$ is the Pochhammer symbol. Similarly, in the case of $\sigma_b\gg\sigma_u,\rho_u\rho_b,d$, the gene is mostly in the bound state and the protein number distribution is negative binomial and given by
\begin{equation}\label{NB2}
p_n = \frac{(s)_n}{n!}p^nq^s.
\end{equation}

Our analytic results can also be used to derive explicit expressions for several other quantities of interest. For example, the steady-state probability that the gene is in the bound state can be recovered from the generating function $f(z)$ at $z = 1$,
\begin{equation}\label{boundprob}
p_{G^*} = \frac{r\mu pq^{-s}K}{(q+\mu)\beta}\hyper(\alpha_1+1,\alpha_2+1;\beta+1;w(1-z_0)).
\end{equation}
In addition, solving Eqs. \eqref{geneqss1} and \eqref{geneqss2} simultaneously for $f'(z)$ and $g'(z)$, we obtain
\begin{equation*}
zf'(z)+g'(z) = \frac{spz}{1-pz}f(z)+\frac{rp}{1-pz}g(z).
\end{equation*}
Substituting $z = 1$ in this equation yields
\begin{equation*}
F'(1) = sBf(1)+rBg(1).
\end{equation*}
where $B = p/q$ is the mean protein burst size. Since $\langle n\rangle = F'(1)$, the steady-state mean of protein number is given by
\begin{equation}\label{mean}
\langle n\rangle = sBp_{G^*}+rBp_G,
\end{equation}
where $p_G$ is the steady-state probability of the gene being in the unbound state. The first term on the right-hand side is the mean protein number $sB$ when the gene is in the bound state multiplied by the corresponding probability $p_{G^*}$ and similarly the second term is the mean protein number $rB$ when the gene is in the unbound state multiplied by the corresponding probability $p_G$. We stress here that the protein mean of the stochastic model given by Eqs. \eqref{boundprob} and \eqref{mean} is not generally the same as that obtained by solving the corresponding deterministic rate equations in steady-state conditions. Similarly, the steady-state second moment of protein number is given by (see Appendix B for details)
\begin{equation}\label{secondmoment}
\begin{split}
\langle n^2\rangle =&\; sB[(1+s)B+1]p_{G^*}+rB[(1+r)B+1]p_G \\
&\; +(1-sB+rB)\left[\frac{\sigma_u}{\sigma_b}p_{G^*}-rBp_G\right].
\end{split}
\end{equation}
This is the sum of three terms: the first term is the second moment of the negative binomial distribution associated with the bound gene state, i.e. the second moment of Eq. \eqref{NB2}, multiplied by the corresponding probability; the second term is the second moment of the negative binomial distribution associated with the unbound gene state, i.e. the second moment of Eq. \eqref{NB1}, multiplied by the corresponding probability; the last term can hence be interpreted as that arising from the difference between the exact probability distribution and a mixture of two negative binomials (this term becomes negligible in the limit of slow gene switching as we show in Section 3.3).

\subsection{Regime of fast gene switching}
\begin{figure}[!htb]
\centerline{\includegraphics[width=0.8\textwidth]{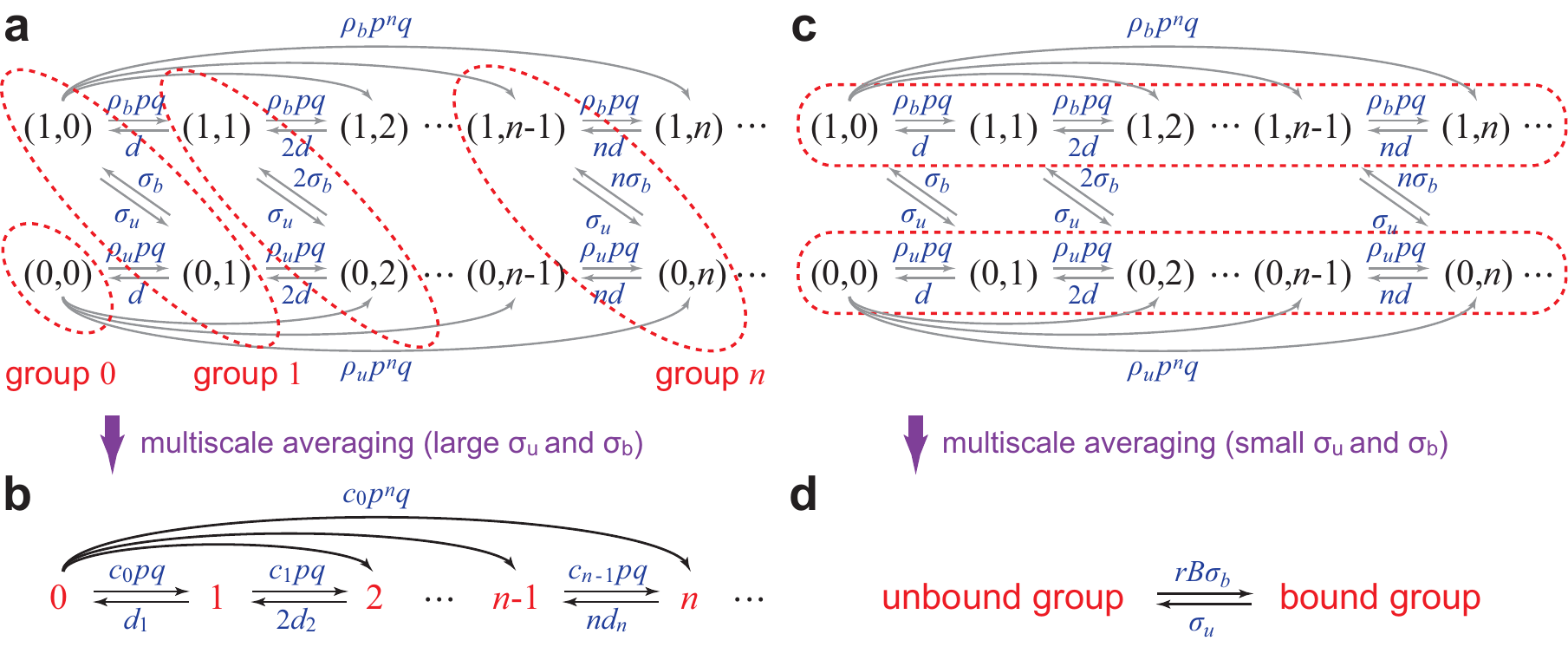}}
\caption{\textbf{Multiscale simplification of the modified Kumar model using averaging method under fast and slow gene switching.} (a) Transition diagram of the modified Kumar model. When the gene switches rapidly between the unbound and bound states, for each $n\geq 1$, the two microstates $(0,n)$ and $(1,n-1)$ can be combined into a group that is labelled by group $n$. (b) Transition diagram of the group dynamics in the limit of fast gene switching. Since group $n$ is composed of two microstates with different protein numbers, the group index $n$ cannot be interpreted as the protein number. (c) Transition diagram of the modified Kumar model. When the gene switches slowly between the unbound and bound states, all unbound microstates $(0,n)$, as well as all bound microstates $(1,n)$, can be combined into a group. (d) Transition diagram of the two-state group dynamics in the limit of slow gene switching.}\label{fast}
\end{figure}

We next focus on two nontrivial special cases. Consider the limiting case when the gene switches rapidly between the unbound and bound states, i.e. $\sigma_u,\sigma_b\gg \rho_u,\rho_b,d$. In this case, the modified Kumar model depicted in Fig. \ref{model}(d) can be further simplified using another classical simplification method of multiscale Markov jump processes called averaging \cite{bo2016multiple, jia2016model}. Since $\sigma_u$ and $\sigma_b$ are large, for any $n\geq 1$, the two microstates $(0,n)$ and $(1,n-1)$ are in rapid equilibrium and thus can be aggregated into a group that is labeled by group $n$, as shown in Fig. \ref{fast}(a). In addition, group 0 is composed of the single microstate $(0,0)$. In this way, the modified Kumar model can be further simplified to the Markovian model illustrated in Fig. \ref{fast}(b), whose state space is given by
\begin{equation*}
\{\textrm{group\;}0, \textrm{group\;}1, \cdots, \textrm{group\;}n, \cdots\}.
\end{equation*}
Here we emphasize that the group index $n$ cannot be interpreted as the protein number. This is because when $n\geq 1$, group $n$ is composed of two microstates with different protein numbers.

The remaining question to address is how to calculate the effective transition rates between two groups. In the fast switching limit, the two microstates $(0,n)$ and $(1,n-1)$ will reach a quasi-steady-state with quasi-steady-state distribution
\begin{equation*}
p^{qss}_{(0,n)} = \frac{\sigma_u}{\sigma_u+n\sigma_b},\;\;\;
p^{qss}_{(1,n-1)} = \frac{n\sigma_b}{\sigma_u+n\sigma_b}.
\end{equation*}
For convenience, we define the effective transcription rate as
\begin{equation}\label{effective}
c_n = \frac{\sigma_u\rho_u+n\sigma_b\rho_b}{\sigma_u+n\sigma_b}
\end{equation}
and the effective protein degradation rate as
\begin{equation}\label{effective2}
d_n = d\left[1-\frac{\sigma_b}{\sigma_u+n\sigma_b}\right].
\end{equation}
It is important to here emphasize that fast switching leads to effective transcription and degradation rates whereas it is customary to write reduced master equations in this parameter regime which only have effective transcription rates; this explains the discrepancies between conventional and exact master equation reduction reported in \cite{holehouse2019revisiting}.  

According to averaging theory \cite{bo2016multiple, jia2016model}, the effective transition rate from group $n$ to group $n+k$ is given by
\begin{equation*}
p^{qss}_{(0,n)}\tilde{q}_{(0,n),(0,n+k)}+p^{qss}_{(1,n-1)}\tilde{q}_{(1,n-1),(1,n-1+k)}
= c_np^kq
\end{equation*}
and the effective transition rate from group $n$ to group $n-1$ is given by
\begin{equation*}
p^{qss}_{(0,n)}\tilde{q}_{(0,n),(0,n-1)}+p^{qss}_{(1,n-1)}\tilde{q}_{(1,n-1),(1,n-2)}
= nd_n.
\end{equation*}
So far, we have obtained all effective transition rates for the group dynamics (Fig. \ref{fast}(b)).

Let $p^{group}_n$ denote the probability of being in group $n$. Then the evolution of the group dynamics is governed by the master equation
\begin{equation*}
\dot p^{group}_n = \sum_{k=1}^{n-1}c_kp^{n-k}qp^{group}_k+(n+1)d_{n+1}p^{group}_{n+1}
-(c_np+nd_n)p^{group}_n.
\end{equation*}
To solve this equation, we notice that it is recursive with respect to the group index $n$. It involves two variables when $n = 1$, three variables when $n = 2$, and so on. At steady-state, solving this master equation by induction leads to
\begin{equation*}
\begin{split}
p^{group}_n &= K\frac{p^n}{n!}
\frac{c_0}{d_1}\cdot\frac{c_1+d_1}{d_2}\cdots\frac{c_{n-1}+(n-1)d_{n-1}}{d_n},
\end{split}
\end{equation*}
where $K$ is a normalization constant. Given that there are $n$ copies of protein in an individual cell, the gene can exist in either microstate $(0,n)$ or microstate $(1,n)$. Since microstate $(0,n)$ is contained in group $n$ and microstate $(1,n)$ is contained in group $n+1$, the steady-state distribution of protein number is given by
\begin{equation}\label{probfast}
\begin{split}
p_n &= p^{group}_np^{qss}_{(0,n)}+p^{group}_{n+1}p^{qss}_{(1,n)} \\
&= K\frac{(p/d)^n}{n!}c_0(c_1+d_1)\cdots(c_{n-1}+(n-1)d_{n-1})
\left[1+\frac{\sigma_bp(c_n+nd_n)}{\sigma_ud}\right].
\end{split}
\end{equation}

In fact, this steady-state protein distribution is exactly the same as that obtained from the generating function given by Eq. \eqref{generating} in the fast switching limit. To see this, we note that when $\sigma_u,\sigma_b\gg \rho_u,\rho_b,d$, the generating function reduces to
\begin{equation*}
F(z) = K(1-pz)^{-s}\left[\hyper(\alpha_1,\alpha_2;\beta;pz)
+\frac{rp}{\beta}\left(1-\frac{s}{r}z\right)\hyper(\alpha_1+1,\alpha_2+1;\beta+1;pz)\right],
\end{equation*}
where
\begin{equation*}
\alpha_1+\alpha_2 = \frac{b}{\mu}-s-1,\;\;\;
\alpha_1\alpha_2 = s+\frac{b(r-s)}{\mu},\;\;\;
\beta = \frac{b}{\mu}.
\end{equation*}
To proceed, we recall the following Euler-Pfaff transformation for hypergeometric functions:
\begin{equation*}
\hyper(\alpha_1,\alpha_2;\beta;z)
= (1-z)^{\beta-\alpha_1-\alpha_2}\hyper(\beta-\alpha_1,\beta-\alpha_2;\beta;z).
\end{equation*}
Using this transformation, the generating function can be simplified as
\begin{equation*}
F(z) = K\bigg[(1-pz)\hyper(\beta-\alpha_1,\beta-\alpha_2;\beta;pz)
+\frac{rp}{\beta}\left(1-\frac{s}{r}z\right)\hyper(\beta-\alpha_1,\beta-\alpha_2;\beta+1;pz)\bigg].
\end{equation*}
Therefore, the steady-state protein distribution can be recovered taking the derivatives of the generating function at zero:
\begin{equation}\label{probgen}
p_n = rK\frac{p^n(\beta-\alpha_1)_n(\beta-\alpha_2)_n}{n!(\beta+1)_n}
\left[\frac{p}{\beta}+\frac{\beta+n}{(\beta-\alpha_1+n-1)(\beta-\alpha_2+n-1)}\right].
\end{equation}
Straightforward computations show that
\begin{equation}\label{crucial1}
\frac{(\beta-\alpha_1+n-1)(\beta-\alpha_2+n-1)}{\beta+n} = \frac{c_n+nd_n}{d}.
\end{equation}
This equality, together with the fact that $r = c_0/d$, shows that
\begin{equation}\label{crucial2}
\frac{r(\beta-\alpha_1)_n(\beta-\alpha_2)_n}{(\beta+1)_n}
= \frac{c_0}{d}\cdot\frac{c_1+d_1}{d}\cdots\frac{c_n+nd_n}{d}.
\end{equation}
Inserting Eqs. \eqref{crucial1} and \eqref{crucial2} into Eq. \eqref{probgen} again yields Eq. \eqref{probfast}.

\subsection{Regime of slow gene switching}
We next consider the limiting case when the gene switches slowly between the unbound and bound states, i.e. $\sigma_u,\sigma_b\ll \rho_u,\rho_b,d$. In this case, the modified Kumar model can also be simplified using the averaging method \cite{bo2016multiple, jia2016model}. Since $\sigma_u$ and $\sigma_b$ are small, all unbound microstates $(0,n)$, as well as all bound microstates $(1,n)$, are in rapid equilibrium and thus can be aggregated into a group, as shown in Fig. \ref{fast}(c). In this way, the modified Kumar model can be further simplified to the Markovian model illustrated in Fig. \ref{fast}(d), which has only two states.

In the slow switching limit, it follows from Eqs. \eqref{NB1} and \eqref{NB2} that all unbound microstates $(0,n)$ will reach a quasi-steady-state with quasi-steady-state distribution
\begin{equation*}
p^{qss}_{(0,n)} = \frac{(r)_n}{n!}p^nq^r,
\end{equation*}
and all bound microstates $(1,n)$ will reach a quasi-steady-state with quasi-steady-state distribution
\begin{equation*}
p^{qss}_{(1,n)} = \frac{(s)_n}{n!}p^nq^s.
\end{equation*}
According to averaging theory \cite{bo2016multiple, jia2016model}, the effective transition rate from the unbound group to the bound group is given by
\begin{equation*}
\sum_{n=1}^\infty p^{qss}_{(0,n)}\tilde{q}_{(0,n),(1,n-1)} = rB\sigma_b
\end{equation*}
and the effective transition rate from the bound group to the unbound group is given by
\begin{equation*}
\sum_{n=0}^\infty p^{qss}_{(1,n)}\tilde{q}_{(1,n),(0,n+1)} = \sigma_u.
\end{equation*}
So far, we have obtained all effective transition rates for the two-state group dynamics (Fig. \ref{fast}(d)).

Let $p_G$ and $p_{G^*}$ denote the steady-state probabilities of being in the unbound and bound groups, respectively. Clearly, we have
\begin{equation*}
p_G = \frac{\sigma_u}{\sigma_u+rB\sigma_b},\;\;\;
p_{G^*} = \frac{rB\sigma_b}{\sigma_u+rB\sigma_b}.
\end{equation*}
Therefore, the steady-state distribution of protein number is given by
\begin{equation*}
\begin{split}
p_n &= p_Gp^{qss}_{(0,n)}+p_{G^*}p^{qss}_{(1,n)} \\
&= \frac{\sigma_u}{\sigma_u+rB\sigma_b}\frac{(r)_n}{n!}p^nq^r
+\frac{rB\sigma_b}{\sigma_u+rB\sigma_b}\frac{(s)_n}{n!}p^nq^s,
\end{split}
\end{equation*}
which is a mixed negative binomial distribution. Since a negative binomial distribution is unimodal, the mixture of two negative binomials can yield a bimodal steady-state protein distribution. It is not hard to show that this steady-state protein distribution is exactly the same as that obtained from the generating function given by Eq. \eqref{generating} in the slow switching limit. The details are omitted here. Note that in the slow switching limit, we have $\sigma_up_{G^*} = rB\sigma_bp_G$, implying that the third term in Eq. \eqref{secondmoment} vanishes leaving only the terms associated with the conditional negative binomials of each gene state.

\subsection{Gene expression noise}
\begin{figure}[!htb]
\centerline{\includegraphics[width=1.0\textwidth]{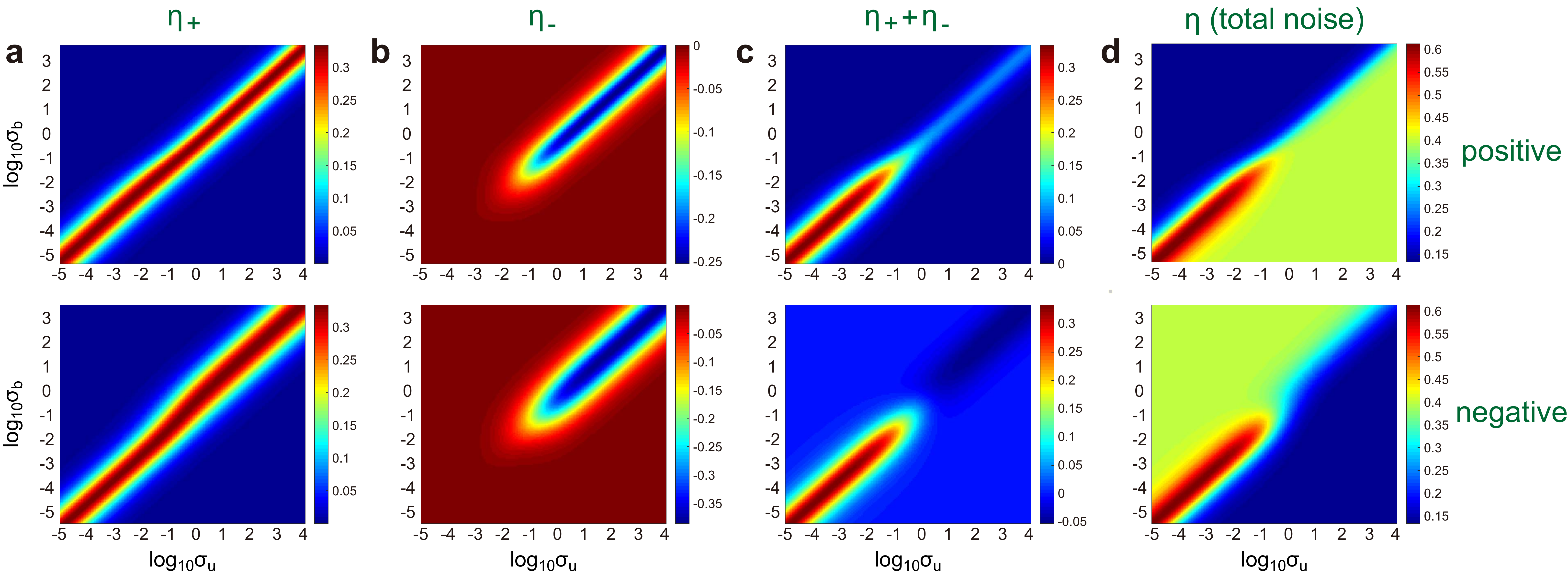}}
\caption{\textbf{Dependence of steady-state protein noise on gene switching rates}. Upper panels are for positive feedback and lower panels for negative feedback. In (a),(b),(c) and (d) we show the dependence of $\eta_+$, $\eta_-$, $\eta_++\eta_-$ and $\eta$, respectively, on the gene switching rates, $\sigma_u$ and $\sigma_b$. In the positive feedback case, the model parameters are $\rho_u = 5, \rho_b = 15, d = 1, p = 0.5$ while in the negative feedback case, the model parameters are $\rho_u = 15, \rho_b = 5, d = 1, p = 0.5$.}\label{noise}
\end{figure}

In single-cell experiments, the size of fluctuations around the protein mean is often measured by the squared coefficient of variation $\eta = \sigma^2/\langle n\rangle^2$, where $\langle n\rangle$ is the mean and $\sigma^2 = \langle n^2\rangle-\langle n\rangle^2$ is the variance. From Eqs. \eqref{mean} and \eqref{secondmoment}, the steady-state variance of protein number is given by
\begin{equation*}
\sigma^2 = (1+B)\langle n\rangle+(s-r)^2B^2p_{G^*}p_G +(1-sB+rB)(Lp_{G^*}-rBp_G).
\end{equation*}
where $L = \sigma_u/\sigma_b$. Therefore, the size of protein number fluctuations can be decomposed into three terms as
\begin{equation*}
\eta = \frac{1+B}{\langle n\rangle}+\eta_++\eta_-,
\end{equation*}
where
\begin{equation}\label{compare1}
\eta_+ = \frac{(s-r)^2B^2p_{G^*}p_G}{(sBp_{G^*}+rBp_G)^2},\;\;\;
\eta_- = \frac{(1-sB+rB)(Lp_{G^*}-rBp_G)}{(sBp_{G^*}+rBp_G)^2}.
\end{equation}
Here $(1+B)/\langle n\rangle$ is the noise due to unregulated bursty gene expression which has been previously derived in the literature \cite{ozbudak2002regulation}. In the presence of autoregulation, two additional terms $\eta_+$ and $\eta_-$ emerge. In Fig. \ref{noise} we show how $\eta_+$, $\eta_-$, their sum $\eta_++\eta_-$, and the total noise $\eta$ vary as a function of the gene switching rates $\sigma_u$ and $\sigma_b$ for positive (upper panels) and negative (lower panels) feedback loops.

Clearly, $\eta_+$ is always positive. Since $0\leq p_{G^*},p_G\leq 1$ and $p_{G^*}+p_G = 1$, the term $\eta_+$ has the following lower and upper bounds:
\begin{equation*}
0 \leq \eta_+ \leq \frac{(s-r)^2}{4sr}.
\end{equation*}
When $L\gg 1$ or $L\ll 1$, we have $p_G \approx 0$ or $p_{G^*} \approx 0$. In this case, the lower bound is achieved and $\eta_+$ vanishes. Moreover, the upper bound is achieved when
\begin{equation*}
p_{G^*} = \frac{s}{s+r}.
\end{equation*}
In this case, the ratio $L$ of $\sigma_u$ and $\sigma_b$ is neither too small nor too large. In other words, in order to obtain a large $\eta_+$, $\log\sigma_b-\log\sigma_u$ must be controlled within a ``narrow" belt (Fig. \ref{noise}(a)).

We next focus on the term $\eta_-$. In the slow switching limit, we have $Lp_{G^*} = rBp_G$ and thus $\eta_-$ vanishes (Fig. \ref{noise}(b)). In the fast switching limit, however, the pair of reversible reactions $G+P\rightleftharpoons G^*$ are in rapid equilibrium and thus the following approximation is appropriate:
\begin{equation*}
\langle n\rangle p_G \approx Lp_{G^*}.
\end{equation*}
This approximation, together with Eq. \eqref{mean}, shows that
\begin{equation}\label{app}
Lp_{G^*}-rBp_G \approx (\langle n\rangle-rB)p_G = (s-r)Bp_{G^*}p_G.
\end{equation}
In addition, we note that since $|s-r|B$, i.e. the absolute difference between the protein mean in the two genetic states, is usually larger than 1 in living cells, then the signs of $1-sB+rB$ and $(r-s)B$ are typically the same. Therefore, in most biologically relevant cases, $\eta_-$ is the product of two terms with different signs and is thus negative in the regime of fast gene switching (Fig. \ref{noise}(b)).

In previous works, it has been shown that for an unregulated gene, the protein noise only contains the Poisson noise and mRNA noise \cite{ozbudak2002regulation}. Therefore, the sum of $\eta_+$ and $\eta_-$ characterizes the contribution of an autoregulatory feedback loop to protein fluctuations. In the slow switching limit, $\eta_-$ vanishes and thus the overall feedback contribution is $\eta_+$. In the fast switching limit, $\eta_-$ is strictly negative and thus the overall feedback contribution $\eta_++\eta_-$ is less than that in the slow switching limit. From Eq. \eqref{app}, in the fast switching limit, $\eta_-$ counteracts most of $\eta_+$ and the remaining term is given by
\begin{equation}\label{compare2}
\eta_++\eta_- \approx \frac{(s-r)Bp_{G^*}p_G}{(sBp_{G^*}+rBp_G)^2},
\end{equation}
which is proportional to $s-r$. Consequently, the overall feedback contribution is positive (negative) for positive (negative) feedback loops. This clearly shows that in the regime of fast gene switching, positive feedback enhances protein noise and negative feedback diminishes it (Fig. \ref{noise}(c)). This is consistent with recent theoretical results obtained in \cite{jia2017stochastic}. In the fast switching limit, comparing Eqs. \eqref{compare1} and \eqref{compare2}, we obtain
\begin{equation*}
\eta_+ \approx (s-r)B(\eta_++\eta_-).
\end{equation*}
Since $|s-r|B$ is the absolute difference between the protein mean in the two genetic states, the overall feedback contribution in the fast switching limit is much smaller than that in the slow switching limit when the protein mean is relatively large (Fig. \ref{noise}(c)).

The total noise $\eta$ is then the superposition of the unregulated contribution $(1+B)/\langle n\rangle$ and the regulated contribution $\eta_++\eta_-$. When $L\gg 1$, the gene is mostly in the unbound state and thus the protein mean has a negative binomial distribution with mean $rB$ and variance $rB(B+1)$. In this case, the total noise is given by
\begin{equation*}
\eta = \frac{rB(B+1)}{(rB)^2} = \frac{B+1}{rB}.
\end{equation*}
Similarly, when $L\ll 1$, the gene is mostly in the bound state and thus the protein mean has a negative binomial distribution with mean $sB$ and variance $sB(B+1)$. In this case, the total noise is given by
\begin{equation*}
\eta = \frac{sB(B+1)}{(sB)^2} = \frac{B+1}{sB}.
\end{equation*}
This explains why the upper-left and lower-right corners in Fig. \ref{noise}(d) have different colors. Due to large feedback contribution, the total noise is also large in the regime of slow gene switching, as can be seen from the lower-left corner in Fig. \ref{noise}(d). In the regime of fast gene switching, feedback regulation gives rise to a weaker enhancement or suppression of the total noise, depending on the sign of the feedback loop.

\subsection{Numerical simulations}
\begin{figure}[!htb]
\centerline{\includegraphics[width=1.0\textwidth]{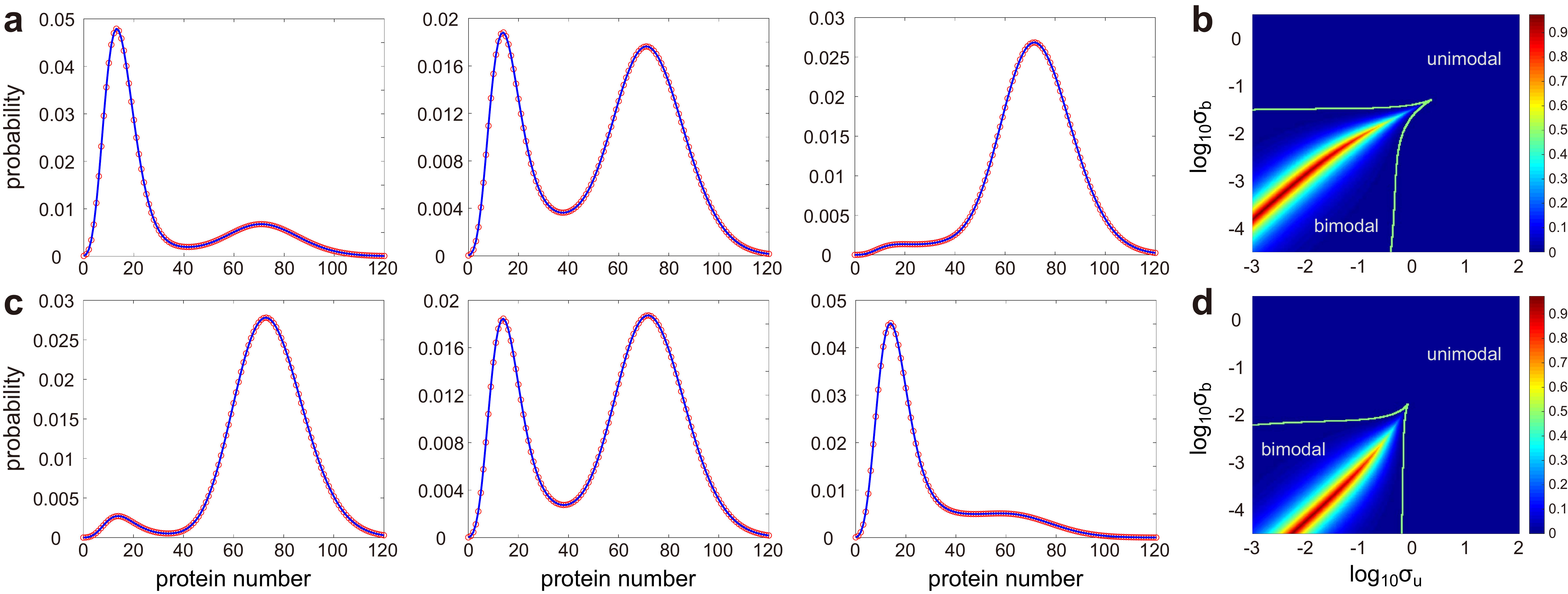}}
\caption{\textbf{Steady-state protein number distribution in the regime of slow gene switching}. (a) Comparison of the analytic solution given by Eq. \eqref{pexactGaussian} (solid blue curve) with the the numerical solution obtained using the SSA (red circles) for positive feedback loops. The model parameters are chosen as $\rho_u = 10, \rho_b = 50, \sigma_u = 0.1, d = 1, p = 0.6$. The feedback strength is chosen as $\sigma_b = 2\times 10^{-3}$ (left), $8\times 10^{-3}$ (middle), and $3.2\times 10^{-2}$ (right). (b) Strength of bistability $\kappa$ defined by Eq. \eqref{strengthbistab} versus the gene switching rates $\sigma_u$ and $\sigma_b$ for positive feedback loops. The model parameters in (b) are chosen to be the same as in (a). (c) Comparison of the analytic solution (solid blue curve) with the SSA (red circles) for negative feedback loops. The model parameters are chosen as $\rho_u = 50, \rho_b = 10, \sigma_u = 0.1, d = 1, p = 0.6$. The feedback strength is chosen as $\sigma_b = 7\times 10^{-5}$ (left), $7\times 10^{-4}$ (middle), and $7\times 10^{-3}$ (right). (d) Strength of bistability versus the gene switching rates for negative feedback loops. The model parameters in (d) are chosen to be the same as in (c). Note that the region designated as bimodal is that satisying the criterion $\kappa > 0$.}\label{simulationslow}
\end{figure}

\begin{figure}[!htb]
\centerline{\includegraphics[width=1.0\textwidth]{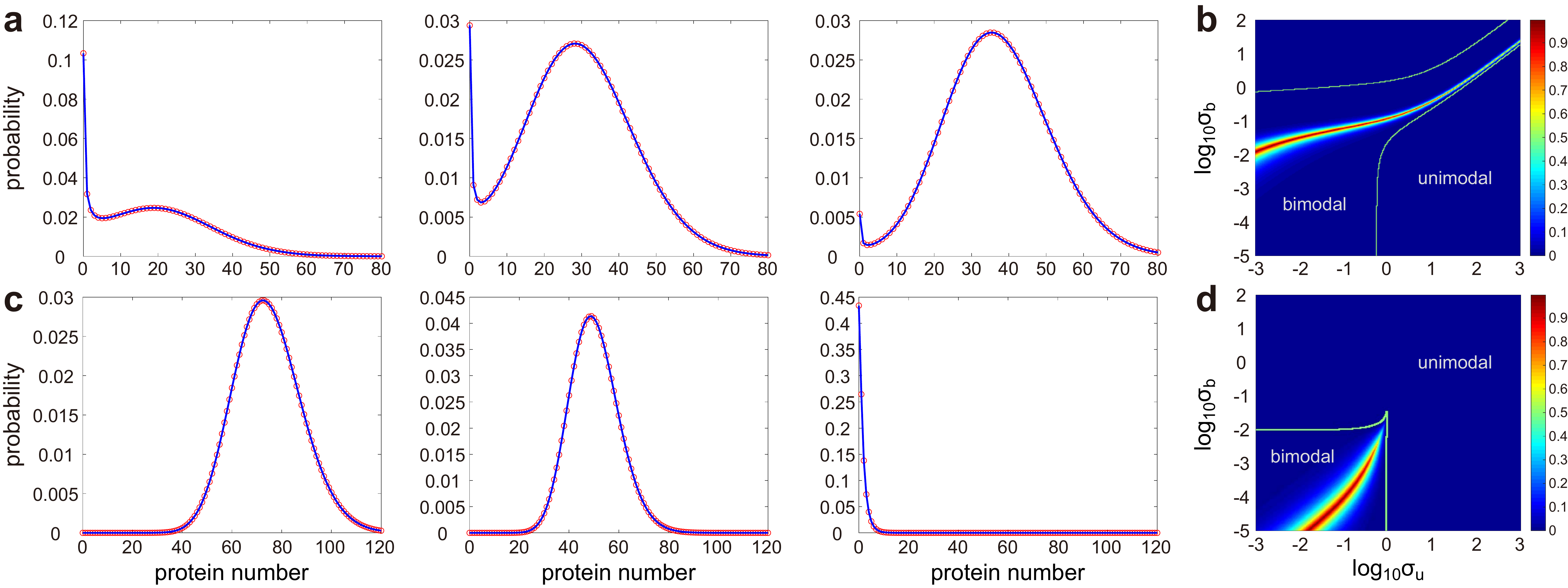}}
\caption{\textbf{Steady-state protein number distribution in the regime of fast gene switching}. (a) Comparison of the analytic solution given by Eq. \eqref{pexactGaussian} (solid blue curve) with the SSA (red circles) for positive feedback loops. The model parameters are chosen as $\rho_u = 0.5, \rho_b = 50, \sigma_u = 1000, d = 1, p = 0.6$. The feedback strength is chosen as $\sigma_b = 10^{1.3}$ (left), $10^{1.37}$ (middle), and $10^{1.44}$ (right). (b) Strength of bistability versus the gene switching rates for positive feedback loops. The model parameters in (b) are chosen to be the same as in (a). (c) Comparison of the analytic solution (solid blue curve) with the SSA (red circles) for negative feedback loops. The model parameters are chosen as $\rho_u = 50, \rho_b = 0.5, \sigma_u = 1000, d = 1, p = 0.6$. The feedback strength is chosen as $\sigma_b = 0.1$ (left), $10$ (middle), and $10^5$ (right). (d) Strength of bistability versus the gene switching rates for negative feedback loops. The model parameters in (d) are chosen to be the same as in (c).}\label{simulationfast}
\end{figure}

To validate our analytic solution given by Eq. \eqref{pexactGaussian}, we compare it with the numerical solution obtained using the stochastic simulation algorithm (SSA) for both positive and negative feedback loops in the regimes of slow (Fig. \ref{simulationslow}(a),(c)) and fast (Fig. \ref{simulationfast}(a),(c)) gene switching. Clearly, the analytic solution coincides perfectly with the SSA. In the regime of slow gene switching, both positive and negative feedback loops can lead to bistable gene expression. In the positive (negative) feedback case, the low expression peak becomes higher as the feedback strength $\sigma_b$ decreases (increases) and the high expression peak becomes higher as $\sigma_b$ increases (decreases) (Fig. \ref{simulationslow}(a),(c)).

To gain a deeper insight into bimodal gene expression, we define the strength of bistability as
\begin{equation}\label{strengthbistab}
\kappa = \frac{h_{\textrm{low}}-h_{\textrm{valley}}}{h_{\textrm{high}}},
\end{equation}
where $h_{\textrm{low}}$ and $h_{\textrm{high}}$ are the heights of the low and high expression peaks, respectively, and $h_{\textrm{valley}}$ is the height of the valley between them. Obviously, $\kappa$ is a quantity between 0 and 1 for bimodal distributions and is set to be 0 for unimodal distributions. In general, to display strong bistability, the following two conditions are necessary: (i) the two peaks should have similar heights and (ii) there should be a deep valley between the two peaks. The former ensures that the time periods spent in the low and high expression states are comparable while the latter guarantees that the two expression levels are distinguishable. Clearly, $\kappa$ is large if the two conditions are both satisfied and $\kappa$ is small if any one of the two conditions is violated. Therefore, $\kappa$ captures both features of bistability and serves as an effective indicator that characterizes its strength.

Using this definition, we investigate how the strength of bistability $\kappa$ varies with the gene switching rates $\sigma_u$ and $\sigma_b$ for positive (Fig. \ref{simulationslow}(b)) and negative (Fig. \ref{simulationslow}(d)) feedback loops when the low expression mode is away from zero. It can be seen that both positive and negative feedback loops can only produce bistability in the regime of slow gene switching. When all model parameters are fixed (except a possible interchange between $\rho_u$ and $\rho_b$), a positive feedback loop requires a larger feedback strength to achieve bistability than a negative feedback loop.

The situation is totally different when the low expression mode lies at zero -- see Fig. \ref{simulationfast}. Interestingly, we find that in this case, a positive feedback loop can produce strong bistability under fast gene switching. This is nontrivial because when gene switching is fast, the effective transcription rate $c_n$ is a Hill-like function with Hill coefficient being equal to 1. In this case, there is no cooperativity in the protein-promoter binding process and the conventional deterministic theory predicts that bistability can never occur (see Appendix C for a proof of this result). However, our stochastic model predicts that a positive feedback loop is capable of bistable behaviour when gene switching is fast even in the absence of cooperative binding. Note that this is not an artefact of the assumptions used to derive the modified Kumar model since simulations verify that it is also a property of the full model. On the other hand, according to our simulations, a negative feedback loop fails to achieve bistable behaviour in the regime of fast gene switching whether the low expression mode is away from zero (Fig. \ref{simulationslow}(d)) or lies at zero (Fig. \ref{simulationfast}(d)). This is consistent with the deterministic prediction.

\section{Reduction of the modified Kumar model to the Grima model}
Consider the limiting case when $\rho_u,\rho_b\rightarrow\infty$ and $p\rightarrow0$, while keeping $\rho_up = \bar\rho_u$ and $\rho_bp = \bar\rho_b$ as constants. Since the mean protein burst size is $B = p/(1-p)$ and the effective protein production rates in the two gene states are $\rho_u B$ and $\rho_b B$, the limits considered above can be interpreted as the limit of negligible mean protein burst size, i.e. $B \rightarrow 0$, taken at constant effective protein production rates. In this case since $q = 1 - p$, we have
\begin{gather*}
\rho_upq \rightarrow \bar\rho_u,\;\;\;\rho_bpq \rightarrow \bar\rho_b,\\
\rho_up^nq \rightarrow 0\;\;\;\rho_bp^nq \rightarrow 0,\;\;\;n\geq 2,
\end{gather*}
and thus the master equation of the modified Kumar model given by Eq. \eqref{masterreduced} reduces to the master equation
\begin{equation*}\left\{
\begin{split}
\dot p_{0,n} &= \bar\rho_up_{0,n-1}+(n+1)dp_{0,n+1}+\sigma_up_{1,n-1}
-(\bar\rho_u+nd+n\sigma_b)p_{0,n},\\ \nonumber
\dot p_{1,n} &= \bar\rho_bp_{1,n-1}+(n+1)dp_{1,n+1}+(n+1)\sigma_bp_{0,n+1}
-(\bar\rho_b+nd+\sigma_u)p_{1,n}.
\end{split}\right.
\end{equation*}
This is exactly the CME of the following chemical reaction system (see Fig. \ref{model}(a) for an illustration):
\begin{gather*}
G+P \xlongrightarrow{\sigma_b} G^*,\;\;\;G^*\xlongrightarrow{\sigma_u} G+P,\\
G\xlongrightarrow{\bar\rho_u} G+P,\;\;\;G^*\xlongrightarrow{\bar\rho_b} G^*+P,\\
P \xlongrightarrow{d} \varnothing.
\end{gather*}
This reaction scheme coincides with the classical model of autoregulatory non-bursty gene networks proposed by Grima et al. \cite{grima2012steady}. The Grima model takes into account changes in protein number during the binding-unbinding process but neglects translational bursting, i.e. it assumes that protein molecules are produced one at a time. Here we have derived the Grima model rigorously as the non-bursty limit of the modified Kumar model which itself is the fast mRNA decaying limit of the full model illustrated in Fig. \ref{model}(c).

To derive the analytic distribution for the Grima model, we normalize all model parameters by the protein degradation rate as
\begin{equation*}
\bar{r} = \frac{\bar\rho_u}{d},\;\;\;\bar{s} = \frac{\bar\rho_b}{d},\;\;\;
\mu = \frac{\sigma_b}{d},\;\;\;b = \frac{\sigma_u}{d}.
\end{equation*}
Recall that when $\alpha_1\rightarrow\infty$ and $z\rightarrow 0$, while keeping $\alpha_1z$ as a constant, the Gaussian hypergeometric function has the following limit
\begin{equation*}
\hyper(\alpha_1,\alpha_2;\beta;z) \rightarrow \confluent(\alpha_2;\beta;\alpha_1z),
\end{equation*}
where $\confluent(\alpha;\beta;z)$ is the confluent hypergeometric function. Taking $\rho_u,\rho_b\rightarrow\infty$ and $p\rightarrow 0$ in Eq. \eqref{generating} and applying the above formula, we obtain
\begin{gather*}
f(z) = \frac{\bar{r}\mu K}{(1+\mu)\beta}e^{\bar{s}z}\confluent(\alpha+1;\beta+1;w(z-z_0)),\\ \nonumber
F(z) = Ke^{\bar{s}z}\bigg[\confluent(\alpha;\beta;w(z-z_0))
+A(1-az)\confluent(\alpha+1;\beta+1;w(z-z_0))\bigg],
\end{gather*}
where
\begin{gather*}
\alpha = \frac{b(\bar r-\bar s)}{\bar r-\bar s-\bar s\mu}-1,\;\;\;
\beta = \frac{b}{1+\mu}+\frac{\bar r\mu}{(1+\mu)^2},\\
A = \frac{\bar{s}-\bar{r}+\bar{r}\mu}{(1+\mu)\beta},\;\;\;
a = \frac{\bar{s}-\bar{r}+\bar{s}\mu}{\bar{s}-\bar{r}+\bar{r}\mu},\;\;\;
w = \frac{\bar{r}}{1+\mu}-\bar{s},\;\;\;z_0 = \frac{1}{1+\mu},
\end{gather*}
and
\begin{equation*}
K = e^{-\bar{s}}[\confluent(\alpha;\beta;w(1-z_0))+A(1-a)\confluent(\alpha+1;\beta+1;w(1-z_0))]^{-1}
\end{equation*}
is a normalization constant that can be determined by solving $F(1) = 1$. This is fully consistent with the results obtained in \cite{grima2012steady}. The steady-state protein distribution for the Grima model can be obtained by taking the derivatives of the generating function $F(z)$ at zero:
\begin{equation*}
\begin{split}
p_n =&\; K\sum_{k=0}^n\frac{(\alpha)_k}{(\beta)_k(1)_k(1)_{n-k}}
\confluent(\alpha+k;\beta+k;-wz_0)w^ks^{n-k} \\
&\; +KA\sum_{k=0}^n\frac{(\alpha+1)_k}{(\beta+1)_k(1)_k(1)_{n-k}}
\confluent(\alpha+1+k;\beta+1+k;-wz_0)w^ks^{n-k} \\
&\; -KAa\sum_{k=0}^{n-1}\frac{(\alpha+1)_k}{(\beta+1)_k(1)_k(1)_{n-1-k}}
\confluent(\alpha+1+k;\beta+1+k;-wz_0)w^ks^{n-1-k}.
\end{split}
\end{equation*}

\section{Comparison of the modified Kumar model with the Kumar model}

\subsection{Regime of slow gene switching}
\begin{figure}[!htb]
\centerline{\includegraphics[width=0.8\textwidth]{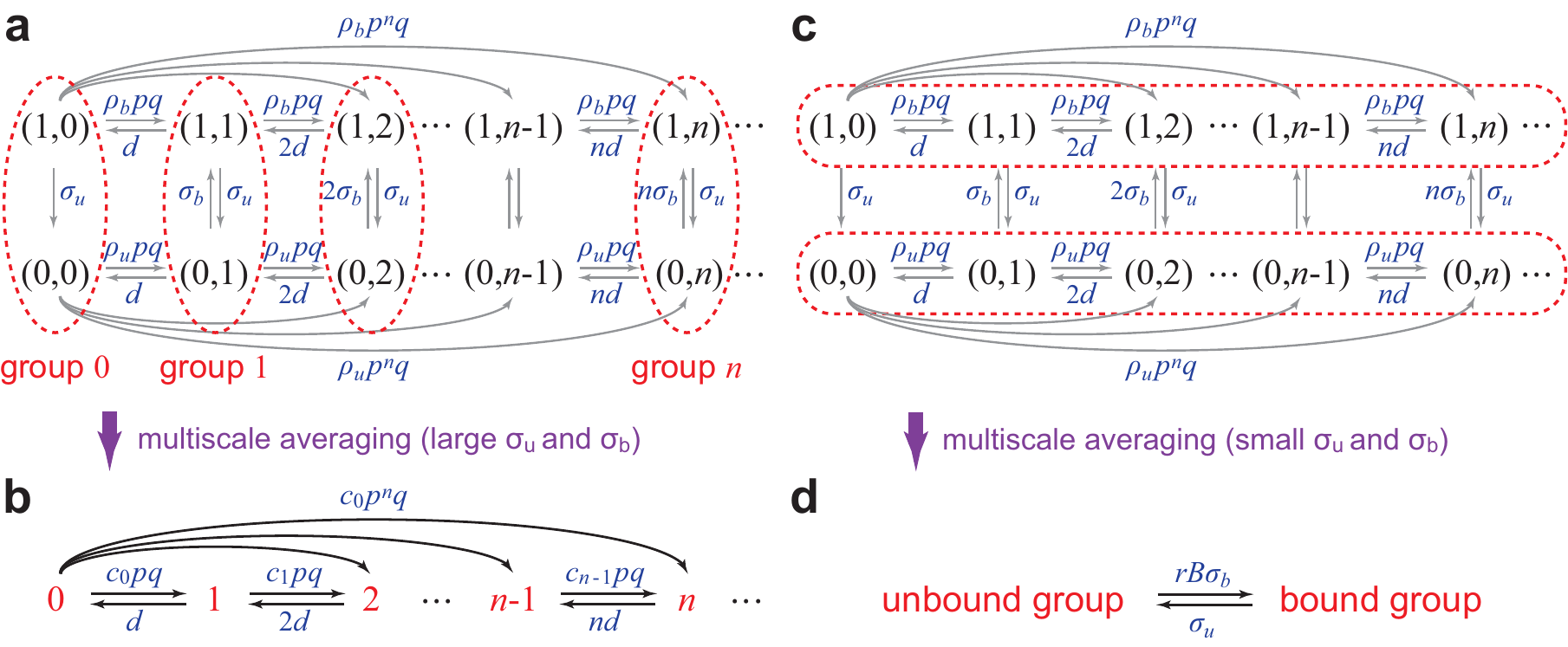}}
\caption{\textbf{Multiscale simplification of the Kumar model using averaging method.} (a) Transition diagram of the Kumar model. When the gene switches rapidly between the unbound and bound states, the two microstates $(0,n)$ and $(1,n)$ can be combined into a group that is labelled by group $n$. (b) Transition diagram of the group dynamics in the limit of fast gene switching. Since group $n$ is composed of two microstates with the same protein number, the group index $n$ can be interpreted as the protein number. (c) Transition diagram of the Kumar model. When the gene switches slowly between the unbound and bound states, all unbound microstates $(0,n)$, as well as all bound microstates $(1,n)$, can be combined into a group. (d) Transition diagram of the two-state group dynamics in the limit of slow gene switching.}\label{fastkumar}
\end{figure}

The modified Kumar model solved in the present paper can also be compared with the classical model of an autoregulatory bursty gene circuit proposed by Kumar et al. \cite{kumar2014exact}:
\begin{gather*}
G+P \xlongrightarrow{\sigma_b} G^*+P,\;\;\;G^*\xlongrightarrow{\sigma_u} G,\\
G\xlongrightarrow{\rho_up^kq} G+kP,\;\;\;G^*\xlongrightarrow{\rho_bp^kq} G^*+kP,\;\;\;k\geq 1,\\
P \xlongrightarrow{d} \varnothing.
\end{gather*}
The Kumar model takes into account translational bursting but neglects protein fluctuations during the binding-unbinding process, i.e. when a protein molecule binds to the promoter, the protein number remains the same. The dynamics of this reaction scheme can be described by the Markovian model illustrated in Fig. \ref{fastkumar}(a).

It has been shown in \cite{kumar2014exact} that the generating function of the Kumar model is given by
\begin{equation*}
F(z) = K(1-pz)^{-s}\hyper(\alpha_1,\alpha_2;\beta;w(z-z_0)),
\end{equation*}
where $K$ is a normalization constant and
\begin{gather}\label{kumar}
\alpha_1+\alpha_2 = \frac{b+r}{1+\mu}-s,\;\;\;
\alpha_1\alpha_2 = \frac{b(r-s)}{1+\mu},\;\;\;
\beta = \frac{b}{1+\mu}+\frac{r\mu p}{(1+\mu)(q+\mu)},\\ \nonumber
w = \frac{(1+\mu)p}{q+\mu},\;\;\;z_0 = \frac{1}{1+\mu}.
\end{gather}
We can see that the generating function of the Kumar model has only one hypergeometric term but, as we proved earlier, the generating function of the modified Kumar model is the superposition of two hypergeometric terms. Comparing Eqs. \eqref{modifiedkumar} and \eqref{kumar}, we find that the parameters $\beta$, $w$, and $z_0$ for the two models are exactly the same, while the parameters $\alpha_1$ and $\alpha_2$ for the two models are different.

\begin{figure}[!htb]
\centerline{\includegraphics[width=1.0\textwidth]{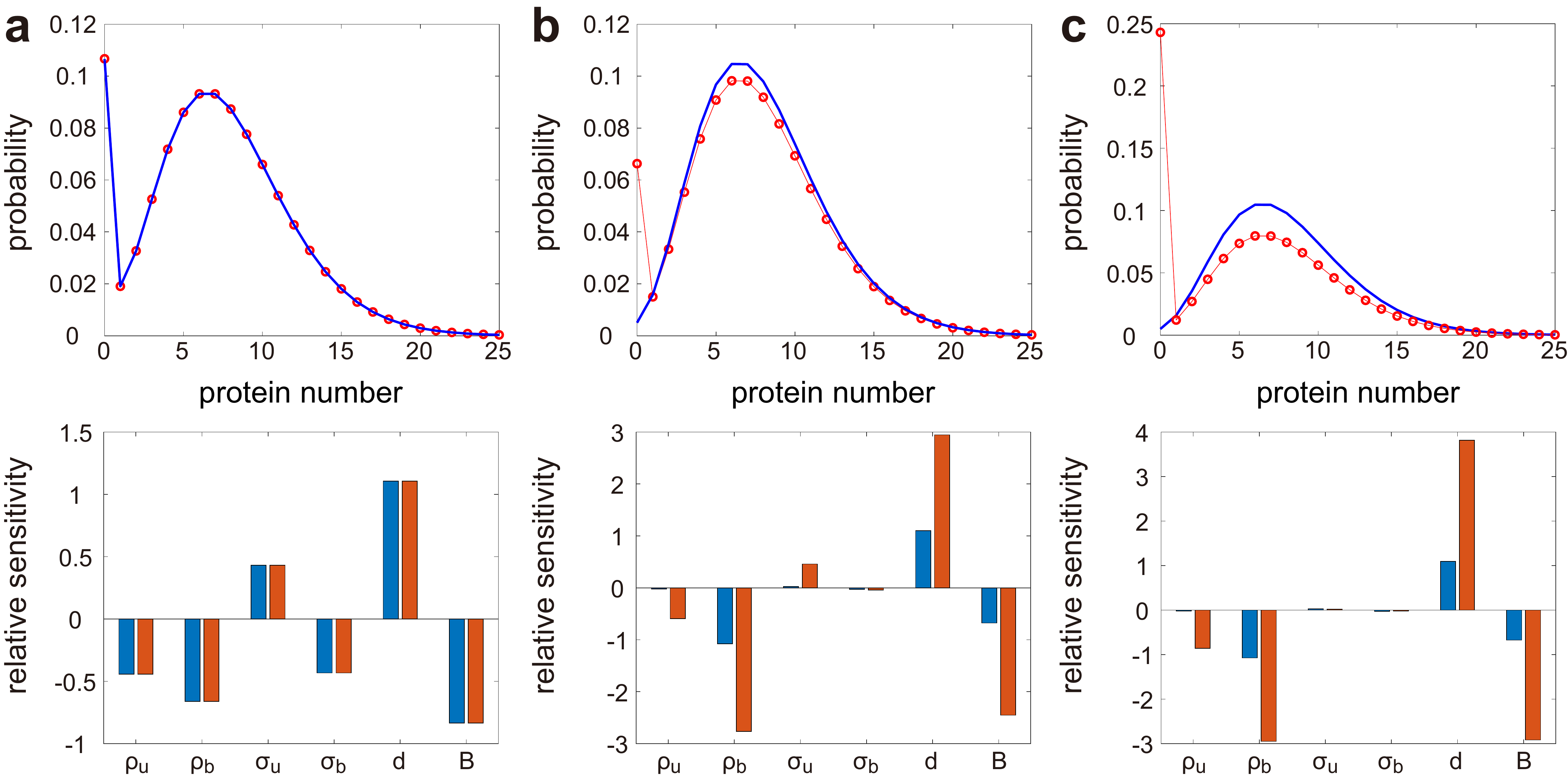}}
\caption{\textbf{Comparison between the steady-state behaviour for the Kumar and modified Kumar models under different gene switching rates}. (a)-(c) Steady-state protein distributions (up) and relative sensitivities of protein noise over various model parameters (down) for the modified Kumar (blue) and Kumar (red) models. (a) Regime of slow gene switching. (b) Regime of moderate gene switching . (c) Regime of fast gene switching. The model parameters are chosen as $\rho_u = 0.1, \rho_b = 8, d = 1, p = 0.5$. The gene switching rates are chosen as $\sigma_u = 10^{-5}, \sigma_b = 8\times 10^{-4}$ in (a), $\sigma_u = 1, \sigma_b = 80$ in (b), and $\sigma_u = 10^5, \sigma_b = 8\times 10^6$ in (c).}\label{change}
\end{figure}

We next focus on the limiting case when the gene switches slowly between the unbound and bound states, i.e. $\sigma_u,\sigma_b\ll\rho_u,\rho_b,d$. Using the averaging method, the Kumar model can be simplified to the same two-state Markovian dynamics (compare Fig. \ref{fastkumar}(c),(d) with Fig. \ref{fast}(c),(d)). Thus, the Kumar model leads to the same steady-state protein distribution as the modified Kumar model in the slow switching limit:
\begin{equation*}
p_n = \frac{\sigma_u}{\sigma_u+rB\sigma_b}\frac{(r)_n}{n!}p^nq^r
+\frac{rB\sigma_b}{\sigma_u+rB\sigma_b}\frac{(s)_n}{n!}p^nq^s.
\end{equation*}
In other words, the Kumar and modified Kumar models share the same steady-state behaviour when gene switching is slow. Fig. \ref{change}(a)-(c) illustrate the steady-state protein distributions for the two models in different regimes of gene switching. It can be seen that the two models agree with each other very well when gene switching is slow, while they disagree, as expected, when gene switching rates are moderate or large.

\subsection{Regime of fast gene switching}
There is still another case when the Kumar model agrees with the modified Kumar model. In the case of $\sigma_u\gg\sigma_b,d$, we have $b\gg 1+\mu$ and thus all the five parameters $\alpha_1,\alpha_2,\beta,w,z_0$ for the two models are the same. Moreover, since $\beta\gg 1$, we have $A\approx 0$ in Eqs. \eqref{generating}-\eqref{modifiedkumar} and thus the generating functions for the two models also coincide with each other. Hence in this case, the two models lead to the same steady-state protein distribution, which can be obtained by taking the derivative of the generating function at $z=0$:
\begin{equation}\label{distfast}
p_n = \sum_{k=0}^n\frac{(\alpha_1)_k(\alpha_2)_k(s)_{n-k}}{(\beta)_k(1)_k(1)_{n-k}}
\frac{\hyper(\alpha_1+k,\alpha_2+k;\beta+k;-wz_0)}{\hyper(\alpha_1,\alpha_2;\beta;w(1-z_0))}
w^kp^{n-k}q^s.
\end{equation}

We next focus on the limiting case when the gene switches rapidly between the unbound and bound states. Since $\sigma_u,\sigma_b\gg \rho_u,\rho_b,d$, the two microstates $(0,n)$ and $(1,n)$ of the Kumar model are in rapid equilibrium and thus can be aggregated into a group, as shown in Fig. \ref{fastkumar}(a). Similarly, using the averaging method, the Kumar model can be simplified to the Markovian model illustrated in Fig. \ref{fastkumar}(b), where $c_n$ is the effective transcription rate defined in Eq. \eqref{effective}.

There are two differences between the Kumar and modified Kumar models in the fast switching limit. For the modified Kumar model, both an effective transcription rate $c_n$ and an effective protein degradation rate $d_n<d$ should be introduced (Fig. \ref{fast}(b)). For the Kumar model, however, only an effective transcription rate $c_n$ should be introduced. In addition, unlike the modified Kumar model, the group index $n$ in the Kumar model can be interpreted as the protein number since each group is composed of two microstates with the same protein number. The steady-state protein distribution for the Kumar model in the fast switching limit is given by \cite{mackey2013dynamic, jia2017stochastic}
\begin{equation}\label{fastkumardist}
p_n = K\frac{(p/d)^n}{n!}c_0(c_1+d)\cdots(c_{n-1}+(n-1)d),
\end{equation}
where $K$ is a normalization constant.

In the fast switching limit, we have seen that the two models lead to the same steady-state protein distribution given by Eq. \eqref{distfast} when $\sigma_b\ll\sigma_u$. However, the two models may deviate significantly from each other when $\sigma_b\gg\sigma_u$ or when $\sigma_b$ and $\sigma_u$ are comparable. In particular, when $\sigma_b\gg\sigma_u$, it follows from Eq. \eqref{NB2} that the steady-state protein distribution for the modified Kumar model is the negative binomial distribution
\begin{equation*}
p_n = \frac{(s)_n}{n!}p^nq^s.
\end{equation*}
Moreover, note that the effective transcription rate $c_n$ has the following approximation when $\sigma_b\gg\sigma_u$:
\begin{equation*}
c_0 = \rho_u,\;\;\;c_n = \rho_b,\;\;\;n\geq 1.
\end{equation*}
Inserting these equations into Eq. \eqref{fastkumardist}, we find that the steady-state protein distribution for the Kumar model is given by a zero-inflated or zero-deflated negative binomial distribution
\begin{equation*}
p_n = \gamma\delta_0(n)+(1-\gamma)\frac{(s)_n}{n!}p^nq^s,
\end{equation*}
where $\delta_0(n)$ is Kronecker's delta which takes the value of 1 when $n = 0$ and takes the value of 0 otherwise, and
\begin{equation}\label{fraction}
\gamma = \frac{s-r}{s+r(q^{-s}-1)}
\end{equation}
is a constant. We make a crucial observation that $\gamma>0$ for a positive feedback loop and $\gamma<0$ for a negative feedback loop. Therefore, when $\sigma_b\gg\sigma_u$, the Kumar model leads to a zero-inflated negative binomial protein distribution in the positive feedback case and leads to a zero-deflated negative binomial protein distribution in the negative feedback case.

\begin{figure}[!htb]
\centerline{\includegraphics[width=1.0\textwidth]{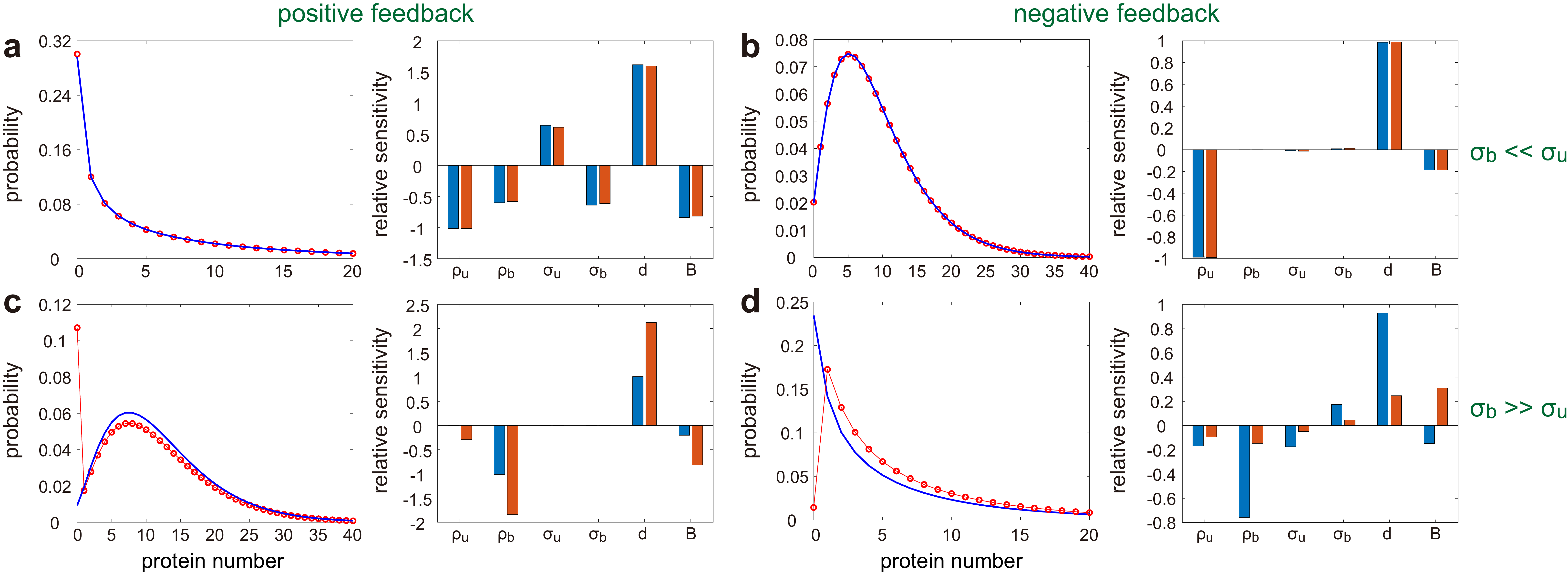}}
\caption{\textbf{Comparison of the steady-state behaviour for the Kumar and modified Kumar models under fast gene switching}. (a),(b) Steady-state protein distributions (left) and relative sensitivities of protein noise to all model parameters (right) for the modified Kumar (blue) and Kumar (red) models when $\sigma_b\ll\sigma_u$. (a) Positive feedback loops. (b) Negative feedback loops. The model parameters are chosen as $\sigma_u = 10^4, \sigma_b = 10^2, d = 1, p = 0.8$. The transcription rates in the two gene states are chosen as $\rho_u = 0.5, \rho_b = 20$ in (a) and $\rho_u = 2.5, \rho_b = 0.1$ in (b). (c),(d) Steady-state protein distributions (left) and relative sensitivities of protein noise to all model parameters (right) for the modified Kumar (blue) and Kumar (red) models when $\sigma_b\gg\sigma_u$. (c) Positive feedback loops. (d) Negative feedback loops. The model parameters are chosen as $\sigma_u = 10^2, \sigma_b = 10^4, d = 1, p = 0.8$. The transcription rates in the two gene states are chosen as $\rho_u = 0.2, \rho_b = 3$ in (c) and $\rho_u = 15, \rho_b = 0.5$ in (d).}\label{comparison}
\end{figure}

To validate our theoretical analysis, we compare the steady-state protein distributions obtained using the SSA for the two models in the regime of fast gene switching (Fig. \ref{comparison}). Clearly, the two models agree with each other perfectly when $\sigma_b\ll\sigma_u$, but they fail as predicted when $\sigma_b\gg\sigma_u$. In the latter case, there is an apparent zero-inflation or zero-deflation in protein number, depending on the type of feedback loop. In the positive (negative) feedback case, the Kumar model has a much higher (lower) probability of having zero protein copy than the modified Kumar model, and the probability of having nonzero protein copies for the Kumar model is lower (higher).

From Eq. \eqref{mean}, we can see that there are two ways of increasing the steady-state protein mean: increasing the transcription rates $\rho_u$ and $\rho_b$ or increasing the mean burst size $B = p/q$. In either case, the term $q^{-s}$ in Eq. \eqref{fraction} becomes larger and thus the constant $\gamma$ becomes closer to zero. When the protein mean is very large, $\gamma$ is almost zero and thus the zero-inflation/deflation effect becomes almost invisible. In other words, the Kumar and modified Kumar models share similar steady-state behaviour when protein mean is large.

\subsection{Sensitivity analysis for protein noise}
Here we investigate the relative sensitivities of protein noise to various model parameters for the Kumar and modified Kumar models. Recall that the relative sensitivity of protein noise with respect to a parameter $s$ is defined as $\Lambda(s) = \partial\log\eta/\partial\log s$, which means that 1\% change in $s$ leads to $\Lambda(s)$\% change in protein noise \cite{ingalls2008sensitivity}.

Since the two models coincide with each other in the regime of slow gene switching, the relative sensitivities of protein noise for the two models should be the same. This prediction is validated by the numerical simulations in Fig. \ref{change}(a)-(c), where the relative sensitivities of protein noise with respect to all model parameters for the two models are compared in different regimes of gene switching. It is clear the two models lead to the same relative sensitivities when gene switching is slow, while they fail to agree when gene switching rates are moderate or large.

In the regime of fast gene switching, we also compare the relative sensitivities of protein noise for the two models in positive and negative feedback cases. The relative sensitivities for the two models agree with each other when protein unbinding is much faster than protein binding (Fig. \ref{comparison}(a),(b)), while they disagree when protein binding is much faster than protein unbinding (Fig. \ref{comparison}(c),(d)). In the positive feedback case, the Kumar model has larger relativity sensitivities than the modified Kumar model. In the negative feedback case, however, the modified Kumar model has larger relativity sensitivities to most of the model parameters compared to the Kumar model. An interesting observation is that the relative sensitivity with respect to a parameter can have a different sign in one model compared to the other model. One such example is the relative sensitivity with respect to $B$ in the negative feedback case with $\sigma_b \gg \sigma_u$; in this case, an increased burst size enlarges protein noise for the Kumar model but diminishes protein noise for the modified Kumar model (Fig. \ref{comparison}(d)).

\section{Conclusions}
In this paper, starting from a stochastic model of an autoregulatory genetic circuit with both mRNA and protein descriptions (the full model), we use the multiscale decimation method to obtain a reduced model with only the protein description, which we refer to as the modified Kumar model. This model takes into account both translational bursting resulting from short-lived mRNA and protein fluctuations stemming from the binding/unbinding reactions with the promoter. Hence it generalizes two classical models of an autoregulatory feedback loop proposed in previous papers: the Kumar model \cite{kumar2014exact}, which takes into account translational bursting but neglects protein fluctuations during the binding-unbinding process, as well as the Grima model \cite{grima2012steady}, which takes into account protein fluctuations during the binding-unbinding process but neglects translational bursting.

We have solved the CME of the modified Kumar model to obtain an exact expression for the steady-state protein number distribution in terms of Gaussian hypergeometric functions. Using the multiscale averaging method, we also obtain the analytic protein distributions in the limits of fast and slow gene switching. In the regime of slow gene switching, the steady-state protein distribution is a mixture of two negative binomials and thus can lead to bistable gene expression for both positive and negative feedback loops. We show that a positive feedback loop requires a larger feedback strength to achieve bistability than a negative feedback loop. Interestingly, we discover that a positive feedback loop can produce noise-induced bistability in the regime of fast gene switching even in the absence of cooperative binding, while a negative feedback loop cannot.

Moreover, we examine steady-state fluctuations in protein abundance in detail. Based on the exact solutions of steady-state protein mean and protein variance, we decompose protein noise, measured by the squared coefficient of variation, into three terms. The first term characterizes fluctuations due to unregulated bursty gene expression which has been reported in previous work \cite{paulsson2000random, ozbudak2002regulation, shahrezaei2008analytical}. The other two terms emerge due to the presence of autoregulation, one is positive and one is negative in most biologically relevant situations, which collectively characterize the overall feedback contribution to protein noise. When gene switching is slow, the positive term is large and the negative terms vanishes. In this situation, the overall feedback contribution is large, which results in large protein noise in the regime of slow gene switching. When gene switching is fast, however, the negative term counteracts most of the positive term and thus the overall feedback contribution is much weaker, with its sign depending on the type of feedback loop.

We further study the relationship between the modified Kumar model and the other two classical stochastic models of autoregulation. In the limit of small mean burst size, we have proved that the modified Kumar model reduces to the Grima model. In addition, we show that the modified Kumar model shares the same steady-state behaviour as the Kumar model under slow gene switching. In the regime of fast gene switching, however, the two models agree with each other when the binding rate of the protein to the promoter is much smaller than the unbinding rate, but deviate significantly from each other when the binding rate is much larger than the unbinding rate. In the latter case, the Kumar model yields an apparent zero-inflation (zero-deflation) in protein number for positive (negative) feedback loops. We have also shown that the relative sensitivities of protein noise with respect to all model parameters, as predicted by the modified Kumar and Kumar models, are typically considerably different in magnitude and in some cases also in sign. These differences as well as the artificial zero-inflation/deflation effect for the Kumar model become increasingly significant as the protein mean decreases, and hence our modified Kumar model provides a more accurate description of fluctuations in the low protein number regime.

In this paper we have obtained the steady-state solution of the modified Kumar model. Potential extensions that are currently under investigation include the derivation of approximate results for the time-dependent protein number distribution for both constant and time-varying transcription rates. 

\section*{Acknowledgements}
C.\ J.\ acknowleges support from startup funds provided by the Beijing Computational Science Research Center. R.\ G.\ acknowledges support from the Leverhulme Trust (RPG-2018-423).

\appendix
\section*{Appendix}

\section{Exact solution of the generating function equations}
In the main text, we have shown that the generating functions of the modified Kumar model satisfy the following system of ODEs:
\begin{equation}\label{twoequations}\left\{
\begin{split}
& [sp(1-z)+b(1-pz)]f-(1-z)(1-pz)f'-\mu(1-pz)g' = 0,\\
& rp(1-z)g-[(1-z)-\mu z](1-pz)g'-bz(1-pz)f = 0.
\end{split}\right.
\end{equation}
Adding the above two equations gives
\begin{equation}\label{firstderivative}
(1-pz)f'-[b(1-pz)+sp]f+(1+\mu)(1-pz)g'-rpg = 0.
\end{equation}
Taking the derivative of this equation yields
\begin{equation}\label{complex}
(1-pz)f''-[b(1-pz)+(1+s)p]f'+bpf+(1+\mu)(1-pz)g''-(1+r+\mu)pg' = 0.
\end{equation}
To proceed, we set
\begin{equation*}
f(z) = (1-pz)^{-s}h(z).
\end{equation*}
It is easy to verify that
\begin{gather}\label{transformderivative}
f' = (1-pz)^{-s}h'+sp(1-pz)^{-s-1}h,\\ \nonumber
f'' = (1-pz)^{-s}h''+2sp(1-pz)^{-s-1}h'+s(s+1)p^2(1-pz)^{-s-2}h.
\end{gather}
Combining Eqs. \eqref{firstderivative} and \eqref{transformderivative} shows that
\begin{equation}\label{1}
\mu g' = -(1-z)(1-pz)^{-s}h'+b(1-pz)^{-s}h.
\end{equation}
Taking the derivative of this equation yields
\begin{equation}\label{2}
\begin{split}
\mu g'' =&\; -(1-z)(1-pz)^{-s}h''-sp(1-z)(1-pz)^{-s-1}h'+(1-pz)^{-s}h'\\
&\; +b(1-pz)^{-\gamma}h'+bsp(1-pz)^{-s-1}h.
\end{split}
\end{equation}
Inserting Eqs. \eqref{transformderivative}, \eqref{1}, and \eqref{2} into Eq. \eqref{complex}, we find that $h$ satisfies the second-order ODE
\begin{equation*}
a(z)h''+b(z)h'-c(z) = 0,
\end{equation*}
where
\begin{equation*}
\begin{split}
a(z) &= (1+\mu)(z-z_0)(1-pz),\\
b(z) &= [1+p+b+\mu+rp-sp]-[(b+r)+(2-s)(1+\mu)]pz,\\
c(z) &= b(1+r-s)p.
\end{split}
\end{equation*}
Note that we have defined
\begin{equation*}
z_0 = \frac{1}{1+\mu}.
\end{equation*}
Since $a(z)$ is a quadratic function of $z$, $b(z)$ is a linear function of $z$, and $c(z)$ is a constant function, the above second-order ODE is a hypergeometric differential equation whose solution is given by
\begin{equation*}
h(z) = \tilde{K}\hyper(\alpha_1+1,\alpha_2+1;\beta+1;w(z-z_0)),
\end{equation*}
where $\tilde{K}$ is a constant and
\begin{gather*}
\alpha_1+\alpha_2 = \frac{b+r}{1+\mu}-s-1,\;\;\;
\alpha_1\alpha_2 = s-\frac{r}{1+\mu}+\frac{b(r-s)}{1+\mu},\\
\beta = \frac{b}{1+\mu}+\frac{r\mu p}{(1+\mu)(q+\mu)},\;\;\;
w = \frac{(1+\mu)p}{q+\mu}.
\end{gather*}
Therefore, we obtain
\begin{equation}\label{f}
f(z) = \tilde{K}(1-pz)^{-s}\hyper(\alpha_1+1,\alpha_2+1;\beta+1;w(z-z_0)).
\end{equation}
Moreover, it follows from Eq. \eqref{firstderivative} that
\begin{equation*}
rpg = (1-pz)f'-[b(1-pz)+sp]f+(1+\mu)(1-pz)g'.
\end{equation*}
This equation, together with the first equation of Eq.  \eqref{twoequations}, implies that
\begin{equation*}
r\mu pg = (1+\mu)(z-z_0)(1-pz)f'-s(1+\mu)p(z-z_0)f+b(1-pz)f.
\end{equation*}
This shows that
\begin{equation}\label{F}
\begin{split}
r\mu pF =&\; r\mu pf+r\mu pg \\
=&\; (1+\mu)(z-z_0)(1-pz)f'-s(1+\mu)p(z-z_0)f+b(1-pz)f+r\mu pf.
\end{split}
\end{equation}
It follows from Eq. \eqref{f} that
\begin{equation}\label{fprime}
f'(z) = \tilde{K}\left[\frac{(\alpha_1+1)(\alpha_2+1)}{\beta+1}w(1-pz)^{-s}H_2
+sp(1-px)^{-s-1}H_1\right],
\end{equation}
where
\begin{equation*}
H_1 = \hyper(\alpha_1+1,\alpha_2+1;\beta+1;w(z-z_0)),\;\;\;
H_2 = \hyper(\alpha_1+2,\alpha_2+2;\beta+2;w(z-z_0)).
\end{equation*}
Inserting Eq. \eqref{f} and Eq. \eqref{fprime} into Eq. \eqref{F} yields
\begin{equation}\label{middle}
\begin{split}
\frac{r\mu p}{\tilde{K}}F =&\; (1+\mu)w\frac{(\alpha_1+1)(\alpha_2+1)}{\beta+1}(z-z_0)(1-pz)^{1-s}H_2\\
&\; +b(1-px)^{1-s}H_1+r\mu p(1-px)^{-s}H_1.
\end{split}
\end{equation}
Recall the following important property of Gaussian hypergeometric functions (see Eq. 15.5.19 in \cite{special})
\begin{equation*}
\begin{split}
&\; z(1-z)(\alpha_1+1)(\alpha_2+1)\hyper(\alpha_1+2,\alpha_2+2;\beta+2;w(z-z_0)) \\
&\; +[\beta-(\alpha_1+\alpha_2+1)z](\beta+1)\hyper(\alpha_1+1,\alpha_2+1;\beta+1;w(z-z_0)) \\
&\; -\beta(\beta+1)\hyper(\alpha_1,\alpha_2;\beta;w(z-z_0)) = 0.
\end{split}
\end{equation*}
This equality implies that
\begin{equation}\label{relation}
w(1+wz_0)\frac{(\alpha_1+1)(\alpha_2+1)}{\beta+1}(z-z_0)(1-pz)H_2
= \beta H_0-[\beta-(\alpha_1+\alpha_2+1)w(z-z_0)]H_1,
\end{equation}
where
\begin{equation*}
H_0 = \hyper(\alpha_1,\alpha_2;\beta;w(z-z_0)).
\end{equation*}
Inserting Eq. \eqref{relation} into Eq. \eqref{middle} yields
\begin{equation*}
\frac{r\mu p}{\tilde{K}}(1-pz)^{s}F = (q+\mu)\beta H_0+(C-Dz)H_1,
\end{equation*}
where
\begin{gather*}
C = b+r\mu p-(q+\mu)[\beta+(\alpha_1+\alpha_2+1)wz_0],\\
D = bp-(q+\mu)(\alpha_1+\alpha_2+1)w.
\end{gather*}
Straightforward calculations show that $C$ and $D$ can be simplified as
\begin{equation*}
C = \frac{(s-r)+r\mu}{(q+\mu)\beta},\;\;\;D = \frac{(s-r)+s\mu}{(q+\mu)\beta}.
\end{equation*}
Therefore, we finally obtain
\begin{equation*}
\begin{split}
F(z) =&\; \frac{\tilde{K}}{r\mu p}(1-pz)^{-s}
\bigg[(q+\mu)\beta \hyper(\alpha_1,\alpha_2;\beta;w(z-z_0)) \\
&\; +(C-Dz)\hyper(\alpha_1+1,\alpha_2+1;\beta+1;w(z-z_0))\bigg].
\end{split}
\end{equation*}

\section{Derivation of the expression for the second moment of protein noise}
Here we shall derive the analytic expression for the steady-state second moment of protein number fluctuations given by Eq. \eqref{secondmoment}. Substituting $z = 1$ in Eq. \eqref{geneqss1}, we obtain
\begin{equation}\label{middle1}
\mu g'(1) = bf(1).
\end{equation}
Taking the derivative of Eq. \eqref{geneqss1} yields
\begin{equation}\label{der1}
\begin{split}
(s+b)pf&-[(1+s)p(1-z)+(1+b)(1-pz)]f'-\mu pg' \\
&+(1-z)(1-pz)f''+\mu(1-pz)g'' = 0.
\end{split}
\end{equation}
Substituting $z = 1$ in this equation gives
\begin{equation}\label{middle2}
\mu g''(1) = -(s+b)Bf(1)+(1+b)f'(1)+\mu Bg'(1).
\end{equation}
Moreover, taking the derivative of Eq. \eqref{geneqss2} yields
\begin{equation}\label{der2}
\begin{split}
b(1-2pz)f+rpg+bz(1-pz)f'&-[(1+r)p(1-z)+(1+\mu)(1-pz)-\mu pz]g' \\
&+[(1-z)-\mu z](1-pz)g'' = 0.
\end{split}
\end{equation}
Substituting $z = 1$ in this equation gives
\begin{equation}\label{middle3}
\mu g''(1) = b(1-B)f(1)+rBg(1)+bf'(1)-[(1+\mu)-\mu B]g'(1).
\end{equation}
Combining Eqs. \eqref{middle2} and \eqref{middle3}, we obtain
\begin{equation*}
f'(1)+(1+\mu)g'(1) = (sB+b)f(1)+rBg(1).
\end{equation*}
Inserting Eq. \eqref{middle1} into this equation yields
\begin{equation}\label{middle4}
\mu f'(1) = (sB\mu-b)f(1)+rB\mu g(1).
\end{equation}
On the other hand, taking the derivative of Eq. \eqref{der1} yields
\begin{equation*}
\begin{split}
2(1+s+b)pf'&-[(2+s)p(1-z)+(2+b)(1-pz)]f''-2\mu pg'' \\
&+(1-z)(1-pz)f'''+\mu(1-pz)g''' = 0.
\end{split}
\end{equation*}
Substituting $z = 1$ in this equation gives
\begin{equation}\label{middle6}
\mu g'''(1) = -2(1+s+b)Bf'(1)+(2+b)f''(1)+2\mu Bg''(1).
\end{equation}
Moreover, taking the derivative of Eq. \eqref{der2} yields
\begin{equation*}
\begin{split}
&-2bpf+2b(1-2pz)f'+2(1+r+\mu)pg'+bz(1-pz)f'' \\
&-[(2+r)p(1-z)+2(1+\mu)(1-pz)-2\mu pz]g''+[(1-z)-\mu z](1-pz)g''' = 0.
\end{split}
\end{equation*}
Substituting $z = 1$ in this equation gives
\begin{equation}\label{middle7}
\mu g'''(1) = -2bBf(1)+2b(1-B)f'(1)+2(1+r+\mu)Bg'(1)+bf''(1)-2[(1+\mu)-\mu B]g''(1).
\end{equation}
Combining Eqs. \eqref{middle6} and \eqref{middle7}, we obtain
\begin{equation}\label{middle8}
f''(1)+(1+\mu)g''(1) = -bBf(1)+[(1+s)B+b]f'(1)+(1+r+\mu)Bg'(1).
\end{equation}
Inserting Eq. \eqref{middle3} into this equation yields
\begin{equation*}
F''(1) = -bf(1)-rBg(1)+(1+s)Bf'(1)+[(1+r)B+1+\mu]g'(1).
\end{equation*}
Moreover, inserting Eqs. \eqref{middle1} and \eqref{middle4} into this equation, we obtain
\begin{equation*}
\mu F''(1) = s(1+s)B^2\mu f(1)+r(1+r)B^2\mu g(1)+(1-sB+rB)[bf(1)-rB\mu g(1)].
\end{equation*}
It follows from Eq. \eqref{mean} that
\begin{equation*}
F'(1) = \langle n\rangle = sBf(1)+rBg(1).
\end{equation*}
Therefore, we have
\begin{equation*}
F''(1)+F'(1) = sB[(1+s)B+1]f(1)+rB[(1+s)B+1]g(1)+(1-sB+rB)\left[\frac{b}{\mu}f(1)-rBg(1)\right].
\end{equation*}
Since $\langle n^2\rangle = F''(1)+F'(1)$, we finally obtain Eq. \eqref{secondmoment} in the main text.

\section{Proof that the deterministic theory only predicts monostability}
In the main text, we have shown that a noisy autoregulatory gene network can perform bistability. Here we show that the traditional deterministic theory fails to predict bistability in any parameter region.

\begin{figure}[!htb]
\centerline{\includegraphics[width=1.0\textwidth]{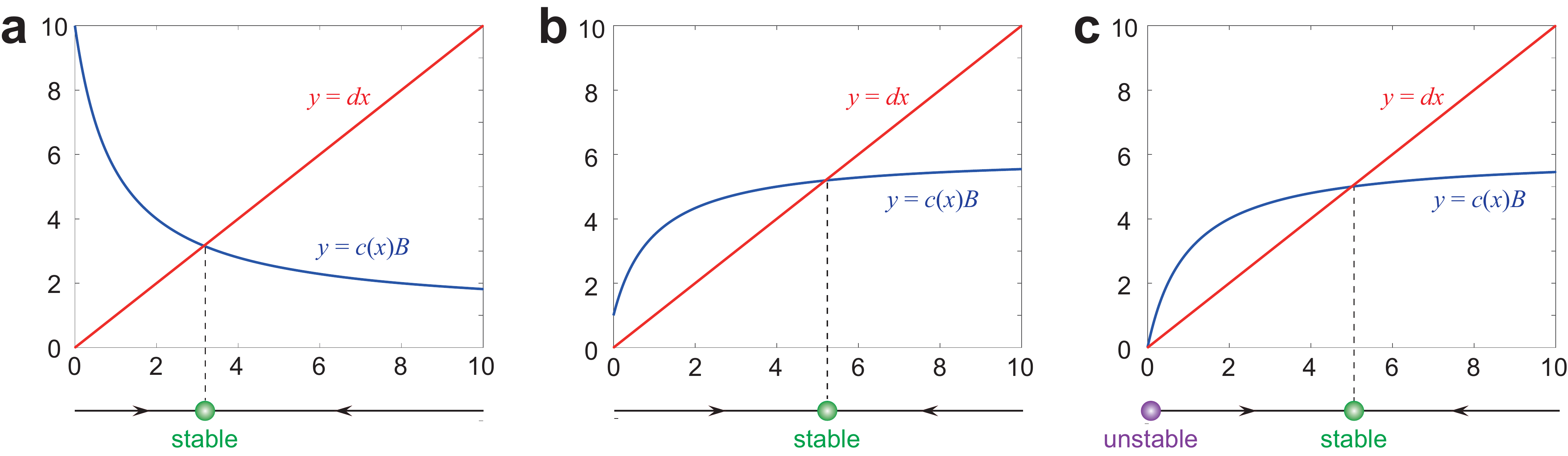}}
\caption{\textbf{Fixed points of the mean-field approximation of the modified Kumar model under fast promoter switching}. (a)-(c) The graphs of the functions $y = c(x)B$ which describes protein synthesis (blue) and $y = dx$ which describes protein degradation (red). The intersections of the two functions give the fixed points of the deterministic model. (a) Negative feedback loops. (b) Positive feedback loops with $\rho_b\neq 0$. (c) Positive feedback loops with $\rho_b = 0$. In (a)-(c), the model parameters are chosen as $\sigma_u = \sigma_b = d = 1, p = 0.5$. The transcription rates in the two gene states are chosen as $\rho_u = 10, \rho_b = 1$ in (a), $\rho_u = 1, \rho_b = 6$ in (b), and $\rho_u = 0, \rho_b = 6$ in (c).}\label{deterministic}
\end{figure}

According to the chemical reaction scheme of the modified Kumar model \eqref{MK} and the law of mass action, the deterministic theory of the modified Kumar model is given by the following set of coupled ODEs:
\begin{equation*}\left\{
\begin{split}
\dot g &= \sigma_bx(1-g)-\sigma_ug,\\
\dot x &= -\sigma_bx(1-g)+\sigma_ug+\rho_uB(1-g)+\rho_bBg-dx,
\end{split}\right.
\end{equation*}
where $g$ is the mean number of genes in the bound state and $x$ is the mean protein number. In steady-state conditions, the gene number equilibrates to
\begin{equation*}
g = \frac{\sigma_bx}{\sigma_u+\sigma_bx}.
\end{equation*}
Substituting this equation into the time-evolution equation for the protein numbers, we obtain
\begin{equation*}
\dot x = c(x)B-dx,
\end{equation*}
where
\begin{equation*}
c(x) = \frac{\rho_u\sigma_u+\rho_b\sigma_bx}{\sigma_u+\sigma_bx}
\end{equation*}
is the effective transcription rate defined in Eq. \eqref{effective}. Fig. \ref{deterministic} shows graphs of the functions $y = c(x)B$ and $y = dx$, whose intersections give the fixed points of the deterministic model. Clearly, there is only one intersection in the negative feedback case, which means that the deterministic model has only one fixed point, which is an attractor (Fig. \ref{deterministic}(a)). In the positive feedback case, however, there are one or two intersections, depending on whether $\rho_u$ vanishes or not. If $\rho_u\neq 0$, the deterministic model has only one fixed point that is an attractor (Fig. \ref{deterministic}(b)). If $\rho_u = 0$, the deterministic model has two fixed points: one is an attractor which is away from zero and the other is an repeller which lies exactly at zero (Fig. \ref{deterministic}(c)). In all cases, there is only one attractor, which implies that the deterministic theory of the modified Kumar model does not allow bistability for any choice of model parameters.

\setlength{\bibsep}{5pt}
\small\bibliographystyle{nature}
\bibliography{Final}

\end{document}